\PassOptionsToPackage{table}{xcolor}
\PassOptionsToPackage{draft}{hyperref}

\documentclass[acmtog]{acmart}
\usepackage{booktabs} 

\setcitestyle{square}
\usepackage{ulem}
\usepackage[ruled]{algorithm2e} 
\usepackage{color,soul}

\SetAlFnt{\small}
\SetAlCapFnt{\small}
\SetAlCapNameFnt{\small}
\SetAlCapHSkip{0pt}
\IncMargin{-\parindent}
\usepackage{subfigure}
\usepackage{caption}
\setcopyright{acmcopyright}
\acmJournal{TOG}
\acmYear{2018} \acmVolume{1} \acmNumber{1} \acmArticle{1} \acmMonth{1} \acmPrice{15.00}\acmDOI{10.1145/3211889}

\setcopyright{usgovmixed}

\acmDOI{0000000.0000000_0}

\usepackage{graphicx}
\usepackage[table]{xcolor}
\usepackage{color}
\usepackage{rotating}
\usepackage{array}
\usepackage{float}
\usepackage{subfigure}
\usepackage{tikz}
\usepackage{pgfplots}
\pgfplotsset{compat=1.9}
\usetikzlibrary{spy, shapes, arrows, matrix, pgfplots.groupplots}
\usetikzlibrary{decorations.text}
\definecolor{myblue}{rgb}{.1,0.3,0.9}

\usepackage{colortbl}
\usepackage{tabularx}
\usepackage{booktabs}

\definecolor{myorchid}{RGB}{10,10,200}
\definecolor{darkgreen}{RGB}{10,150,10}
\definecolor{darkred}{RGB}{100,150,20}
\definecolor{YellowOrange}{RGB}{255,127,39}
\newcommand{\ow}[1] {\emph{\textcolor{myorchid}{OW: #1}}}

\newcommand{\ct}[1]{\emph{\textcolor{darkred}{CT: #1}}}

\newcommand{\ve}[1]{\mbox{{\bf #1}}}
\renewcommand{\deg}{$^{\circ}$~}

\usepackage{booktabs}
\usepackage{xfrac}
\usepackage{tablefootnote}
\usepackage{colortbl}
\usepackage{tabularx}
\definecolor{rowblue}{RGB}{220,230,240}

\usepackage{gensymb}
\usepackage{hyperref}

\renewcommand{\paragraph}[1]{\vspace{8px} \noindent \textbf{#1} \ \ }
\newcommand{\review}[1]{#1}

\newcommand{\tog}[1]{{#1}}
\newcommand{\togb}[1]{\textcolor{red}{#1}}

\begin{document}
\renewcommand\togb{}
	
\title{Joint Stabilization and Direction of 360\deg Videos}

\author{Chengzhou Tang}
\affiliation{%
	\institution{Simon Fraser University}
}
\author{Oliver Wang}
\affiliation{%
	\institution{Adobe Systems Inc}
}
\author{Feng Liu}
\affiliation{%
	\institution{Portland State University}
}
\author{Ping Tan} 
\affiliation{%
	\institution{Simon Fraser University}
}
\renewcommand\shortauthors{Tang, C. et al}









\begin{abstract}
360\deg video provides an immersive experience for viewers, allowing them to freely explore the world by turning their head.
However, creating high-quality 360\deg video content can be challenging, as viewers may miss important events by looking in the wrong direction, or they may see things that ruin the immersion, such as stitching artifacts and the film crew.
We take advantage of the fact that not all directions are equally likely to be observed; most viewers are more likely to see content located at ``true north'', i.e. in front of them, due to ergonomic constraints.
We therefore propose 360\deg video direction, where the video is jointly optimized to orient important events to the front of the viewer and visual clutter behind them, while producing smooth camera motion.
Unlike traditional video, viewers can still explore the space as desired, but with the knowledge that the most important content is likely to be in front of them.
Constraints can be user guided, either added directly on the equirectangular projection or by recording ``guidance'' viewing directions while watching the video in a VR headset, or automatically computed, such as via visual saliency or forward motion direction.
\tog{
To accomplish this, we propose a new motion estimation technique specifically designed for 360\deg video which outperforms the commonly used 5-point algorithm on wide angle video.
We additionally formulate the direction problem as an optimization where a novel parametrization of spherical warping allows us to correct for some degree of parallax effects. 
We compare our approach to recent methods that address stabilization-only and converting 360\deg video to narrow field-of-view video. 
}
Our pipeline can also enable the viewing of wide angle non-360\deg footage in a spherical 360\deg space, giving an immersive ``virtual cinema'' experience for a wide range of existing content filmed with first-person cameras.


\end{abstract}

%
%


%
%



\maketitle
\section{Introduction}
VR headsets are rapidly gaining in popularity, and one of the most common use cases is viewing 360$\degree$ videos, which provide added immersion due to the ability of the viewer to explore a wider field of view than traditional videos.
However, this freedom introduces a number of challenges for content creators and viewers alike; viewers can miss important events by looking in the wrong direction, or they can see things that break immersion, such as stitching artifacts or the camera crew.


 \begin{figure*}[t!]
	\includegraphics[width=1\linewidth]{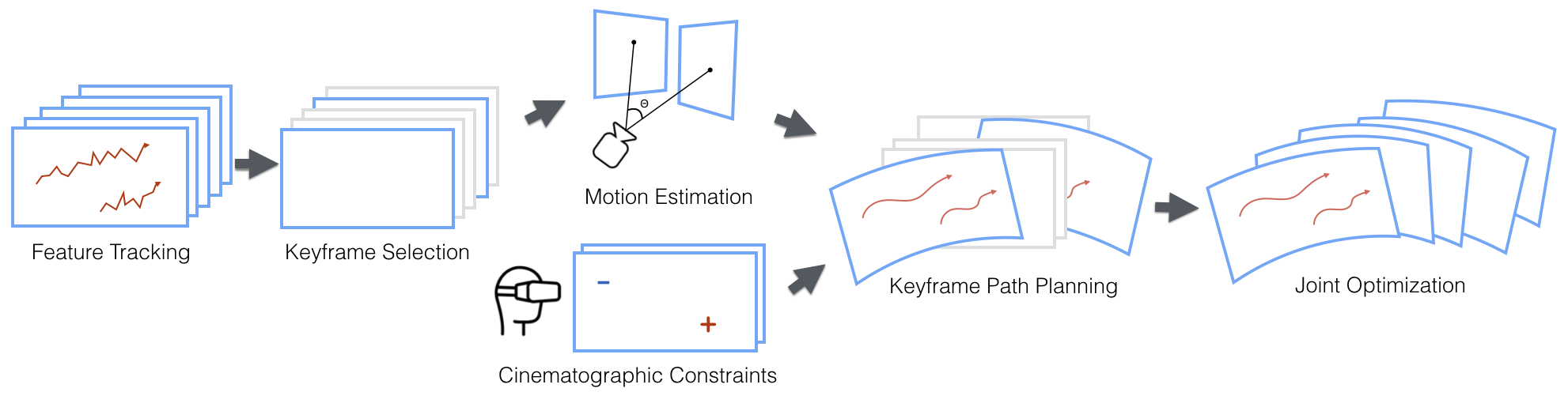}
	\caption{\label{fig:overview} Our method follows the above pipeline; features are tracked across all frames, after which keyframes are selected. A novel 3D camera motion estimation is combined with user-guided viewing constraints in the path planning step to produce a set of aligned keyframes, and the results are finally interpolated onto all frames of the video. 360\deg frames are visualized as rectangles.}
\end{figure*}

\togb{
In this work, we address two important aspects of 360 video creation; direction of shots to draw the viewers' attention to desired regions, and smooth, intentional camera trajectories.
Both of these parts are crucial in 360\deg video; viewer freedom makes direction challenging, and unstable motion (especially in the peripheral vision) can be disorienting, causing confusion and even nausea~\cite{Jerald:2015}.
While traditional cinematography refers to the decisions made \emph{during} filming, 360\deg video is particularly well suited to the task of modifying the camera direction in post (also known as re-cinematography~\cite{gleicher2008re}), as all viewing directions are recorded at capture time, giving us greater control in post.
}


\tog{
We therefore propose a method that uses direction constraints~\cite{gandhi2014multi} (e.g., where to look, where not to look), that try to keep desirable content in the viewable space near true-north, and undesirable content largely behind the user.
These are jointly optimized with smoothness constraints that reduce camera shake and rapid rotations, such as those caused by hand held cameras, motorized gimbals, or inconsistent direction constraints.}
We allow an editor to manually define desirable and undesirable regions in the video, as well as the ability to use automatically derived constraints such as saliency maps, forward motion, or stitching regions for known camera configurations.
In the case of manual constraints, editors can either directly draw on the equirectangular projection, or alternately we propose a new type of interaction where the editor views the content in a VR headset as a ``guide'', and their viewing path is recorded and used as a constraint in the joint optimization.

\togb{
In summary, we propose a solution for joint stabilization and direction of 360\deg videos, where undesirable camera motions (e.g., shake and rapid rotations) are removed while following a smooth and directed camera path.
}
Our solution also works for wide-angle videos, enabling ``virtual cinema'' viewing in VR for a large library of existing footage. 
To achieve these goals, we present the following technical contributions:
\begin{itemize}
   \item A motion estimation algorithm based on non-linear optimization which performs better than widely used five-point algorithm on 360\deg and wide-angle videos.
   \item A 3D spherical warping model derived from our motion estimation that and handles both rotation and translation which allows more control than the recently proposed method~\cite{kopf2016360}.
   \item A unified framework to define and add constraints from different sources on the resulting 360\deg video, including a new VR interface and automatic motion constraints.
\end{itemize}
To validate these contributions, we make both qualitative and quantitative comparisons and conduct user study to show that in our directed results, viewers are much more likely to observe the desired parts of the sequence marked with positive constraints. 

%
\section{Related Work}
360\deg video can now be captured by both consumer handheld cameras, such as Ricoh Theta, Nikon KeyMission, and Kodak Pixpro SP360, and professional camera array systems~\cite{Anderson:2016,Lee:2016:ROS}.
The captured 360\deg video often needs to be post-processed to deliver a pleasant viewing experience. 
We briefly review relevant prior research. 

Video stabilization is a common approach to improve the camera motion of a video. 
Common ways to stabilize video involve tracking features over time and computing an image warp sequence that smooths feature trajectories and thus the apparent motion of the video.
This can be achieved by applying a sequence of homographies or other 2D transformations that compensate the motion~\cite{Chen:2008:CIB,Matsushita:2006,lee2009video}, by using a grid of homographies~\cite{liu2013bundled} for robustness, by using  projective reconstruction~\cite{Goldstein:2012}, on a low dimensional feature subspace~\cite{liu2011subspace}, as used in Adobe After Effects,  or by fitting smooth cropping windows that minimize first and second order motion~\cite{grundmann2011auto}, as used on YouTube.
Alternate approaches have proposed building a full 3D reconstruction of the scene, which can be used to synthesize a smooth virtual camera trajectory~\cite{Buehler:2001:NIR,liu2009content,kopf2014first}.


Recently, Kopf~\shortcite{kopf2016360} presented an extension of video stabilization to 360\deg videos.
This approach computes a 3D geometric relationship between keyframes via the 5-point algorithm~\cite{li2006five}, and smoothly interpolates keyframes using a deformable model.
We use a similar approach with a few significant modifications.
Whereas Kopf~\shortcite{kopf2016360} is largely used to compute a ``total'' stabilization (where fixed scene points will remain stationary throughout the video), we go beyond stabilization and combine artistic and \emph{smoothness} constraints to produce an easy to watch 360\deg video with directed camera motion.
One requirement to support this goal is that we need a full 3D motion rotation and translation estimation per frame.
To achieve this, we introduce a new method for estimating rotation and translation on a sphere that is more robust than the 5-point algorithm.



Our work is inspired by previous work on re-cinematography~\cite{gleicher2008re,gandhi2014multi}, where casually captured video is improved by integrating high-level content driven constraints and low-level camera motion stabilization.
We extend this notion to 360\deg video, which benefits from all viewing angles being captured during filming, so cropping is not necessary, freeing our method from the trade off between guiding the viewer's attention and preserving video content.
Similar path planning in traditional video has also been used for retargeting~\cite{Jain:2015,wang2009motion}, which distorts or crops off less important content to fit the video into a different aspect ratio other than originally intended.

The Pano2Vid work by Su \emph{et al.}~\shortcite{supano2vid} performs a related, but different task of automatically producing a narrow field-of-view video from a 360\deg video.
A recent follow-up work extended this to optimize for zoom as well~\cite{DBLP:journals/corr/SuG17}.
These approaches are complementary to ours; in our work, we combine viewing constraints with motion estimation to compute smooth directed camera paths for 360\deg viewing.
Pano2Vid presents a learning based saliency computation, and computes a shortest path on these values.
As it does not perform any motion estimation, it cannot be used to stabilize camera motion in shaky videos. 
We show that we can use the output from Pano2Vid as automatic saliency constraints for our method, which generates smooth camera paths where important content is placed in front of the viewers.


One of our main assumptions is that by rotating important objects in front of viewers, VR viewing becomes a more enjoyable experience.
This is in some sense, reducing some control of the viewer to freely explore the space, as some rotations will be out of the control of the viewer.
Whether this kind of motion is ``allowable'' is an open topic in VR film making, but we note that traditional wide angle video also started with static shots before filmmakers learned how to use camera motion as an artistic tool without disorienting the viewer/

In this domain, work by Sun et al.~\shortcite{sun2016mapping} has shown that it is in fact possible to separate the viewer's real world head motion from the perceived motion in the VR space in order to map virtual spaces into physical constraints such as room sizes without causing confusion.
Sitzmann et al.~\shortcite{8269807} also conducted user studies and concluded that the faster user attention gets directed towards the salient regions, the more concentrated their attention is.




%
\section{Method}
Figure~\ref{fig:overview} provides an overview of the approach.
We first estimate the existing motion between keyframes using feature tracking and a novel pairwise motion estimation formulation for spherical and wide angle video. 
We then solve a joint optimization on the keyframes that enforces smoothness in the warped video sequence and a set of user-provided (or automatic) path planning constraints.
Finally, we smoothly interpolate the motion between keyframes to produce the final output video.
We now discuss each step in more detail.

\label{sec:stab}

\subsection{Feature Tracking and Keyframe Selection}
\label{sec:feature}
Similar to prior work~\cite{kopf2016360}, for 360\deg videos we remap the equirectangular image to a cube map and track feature points independently on each cube face using KLT feature tracking~\cite{klt}.
For wide-angle videos, we perform the tracking directly on the video frames. 
This is the only stage of the process that is different between 360\deg and traditional wide-angle video, after tracking we project feature points onto a sphere and treat both types of videos identically. 
In the following, we will use unit vector $\ve{p}$ to denote the projected points on a sphere.

Similar to other approaches, we select feature points on keyframes and track them through the video~\cite{kopf2014first,kopf2016360}.
The first and last frames are selected as keyframes, and we create new keyframes every time the percentage of successful tracked features drops to 60\% of the number of features initially detected.
Finally, we only select feature points that are more than 2\deg away from any previously selected feature points.

\tog{
After feature tracking, we have a set of $m$ feature trajectories $T=\{T_i|{i=1\cdots m}\}$ through the video, where each trajectory $T_{i}$ is a list of points from several continuous frames:
\begin{equation}
T_{i}=\{\ve{p}^{i}_j|j=s_i\cdots e_i\},
\label{eq:ft}
\end{equation}
where $s_i$ is the starting frame (which is always a keyframe), and $e_i$ is the last keyframe where the point was successfully tracked.
}



\subsection{Rotation and Translation Estimation}
\label{sec:motionestimation}


After collecting feature tracks, we estimate the relative 3D rotation and translation between neighboring pairs of keyframes.
A common solution is to use the 5-point algorithm \cite{5PT}, to first estimate the essential matrix, decompose it into a rotation matrix $\ve{R}$ and a translation direction vector $\ve{t}$, and then improve the estimated motion by iterative refinement~\cite{Triggs2000}.
With this approach, the final quality relies on the accuracy of the essential matrix estimation as well as the motion decomposition. 
It is well known both of these steps are highly dependent on the quality of the camera calibration~\cite{Stewenius2005}, feature trajectories, and global shutter camera~\cite{DAI2016}.
As a result, prior 360\deg stabilization work only uses the estimated rotation between neighboring keyframe pairs, and discards the 3D translation, which can be unreliable.
The final stabilization is then performed by smoothing feature trajectories directly on the spherical image~\cite{kopf2016360}. 

We propose a new motion estimation method that is robust to poorly calibrated cameras, rolling shutter effects, and errors in the feature trajectories, and yields 3D rotation and translation direction estimates which are accurate enough to be directly used to stabilize the footage, as well as to provide direction such as a forward motion constraint (discussed later in Sec.~\ref{sec:pathplanning}).
Both the 5-point algorithm and our method are two view motion estimation methods, where translation can only be computed up to an unknown scale.
Therefore, we use the standard definition for translation $\ve{t}$ as a $3\times1$ unit vector~\cite{Hartley2004}, representing the translation \emph{direction}.

\begin{figure}
	\centering
	\subfigure[(a) Rotation]{\label{fig:a}\includegraphics[width=0.32\linewidth,page=1]{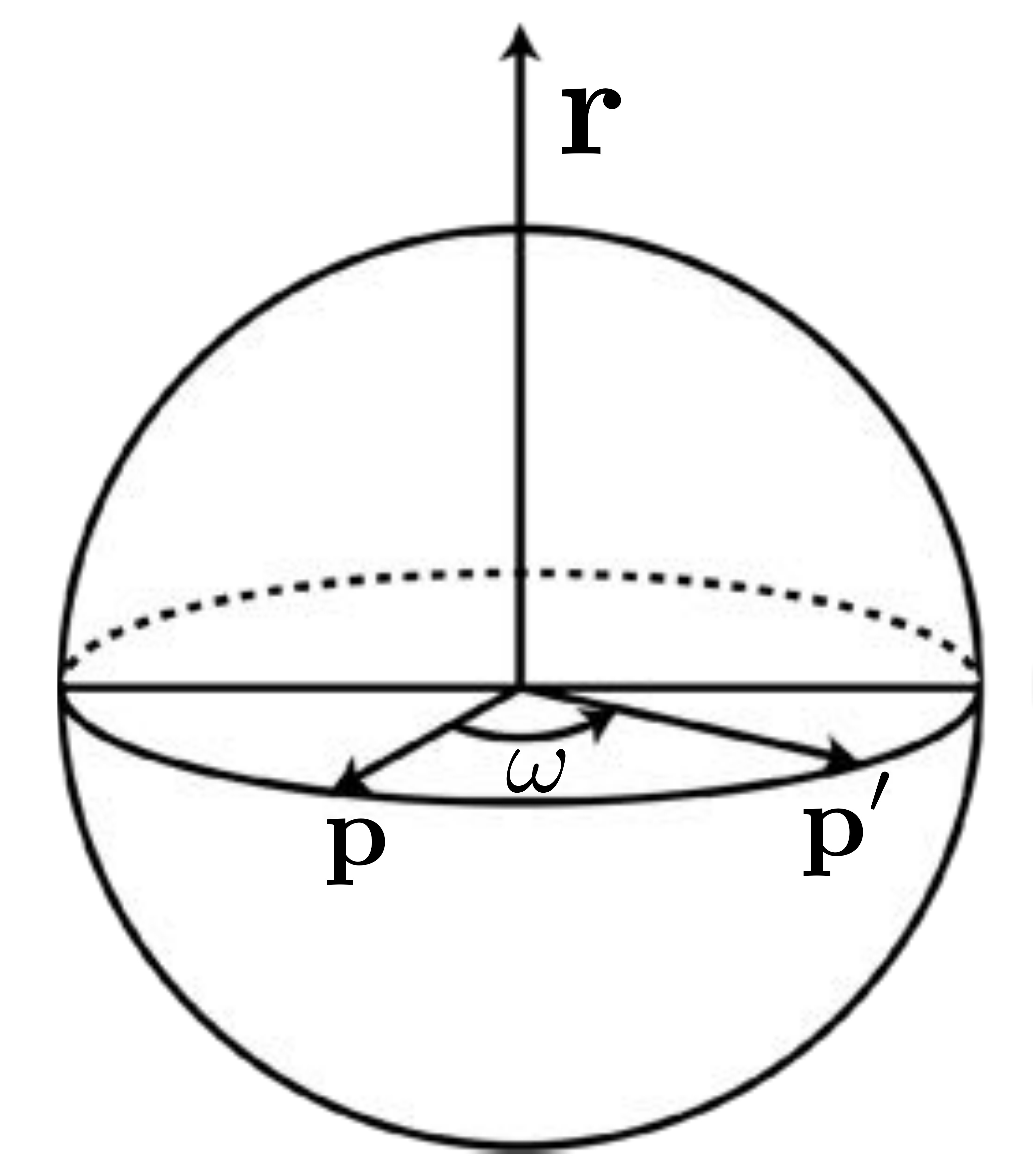}}
	\subfigure[(b) Translation]{\label{fig:b}\includegraphics[width=0.32\linewidth,page=2]{figures/RT.pdf}}
	\subfigure[(c) Combined]{\label{fig:c}\includegraphics[width=0.32\linewidth,page=3]{figures/RT.pdf}}
	\caption{\label{fig:rottrans}The motion of a point ($\ve{p}$ to $\ve{p}'$) on a sphere, when (a) induced by camera rotation around the axis $\ve{r}$, (b) induced by a camera translation in direction $\ve{t}$, and (c), both together. We estimate $\ve{t}$, $\ve{R}$, and per-feature $\theta$s by minimizing distances $E_M$ for all tracked feature locations $\ve{q}$}
\end{figure}

\paragraph{Direct Spherical Motion Estimation}
We first consider the relationship between feature correspondences and the 3D transformation between two frames, a reference frame (with identity rotation and zero translation), and second frame, with rotation \ve{R} and translation \ve{t}.
Figure~\ref{fig:rottrans} illustrates the motion of a feature correspondence $\ve{p}\rightarrow\ve{p}'$ as a function of both 3D rotation $\ve{R}$ and translation $\ve{t}$.
To gain a better understanding let's consider the pure rotational and pure translation cases independently. 
We use Rodrigues' formula~\cite{Gray1980} to represent rotation:
\begin{equation}\label{eq:rotation}
\mathbb{R}(\ve{r},\omega)=\ve{I}+\sin(\omega)\ve{r}_{\times}+(1-\cos(\omega))\ve{r}\otimes\ve{r},
\end{equation}
where $\ve{r}$ is the rotation axis, $\omega$ is the rotation angle, $\ve{r}_{\times}$ is the cross product matrix of $\ve{r}$ and $\otimes$ is the tensor product.

When the camera undergoes a rotation, every feature $\ve{p}$ on the sphere rotates around the same axis $\ve{r}$ with the same angle $\omega$ independent of its depth, i.e.\ $\ve{p}'=\mathbb{R}(\ve{r},\omega)\ve{p}$.
Since every point rotates around the same axis by the same angle we simplify $\mathbb{R}(\ve{r},\omega)$ as a global rotation $\ve{R}$.

When translating, each $\ve{p}$ rotates around a \emph{feature dependent} axis $\ve{t}_{\times}\ve{p}$ by an angle $\theta$ (which is a function of point depth), towards the intersection of $\ve{t}$ and the sphere, i.e.\ $\ve{p}'=\mathbb{R}(\ve{t}_{\times}\ve{p},\theta)\ve{p}$, where $\ve{p}$, $\ve{t}$ and $\ve{p}'$ are in the same plane, and $\ve{t}_{\times}\ve{p}$ is the normal vector of this plane, i.e.\ $(\ve{p}')^{\top}\ve{t}_{\times}\ve{p}=0$. 

Without loss of generality, we define the camera motion $M=[\ve{R},\ve{t}]$ as a rotation $\ve{R}$ followed by a translation $\ve{t}$.
To account for noise in the feature tracks, we estimate the rotation $\ve{R}$, translation $\ve{t}$ and the feature dependent angle $\theta$ by minimizing the following error where $\ve{q}$ is the tracked feature location:
\begin{equation}\label{eq:motionerror0}
E_{M}(\ve{R},\ve{t},\theta)=\|\ve{q}-\ve{p}'\|_{2}=\|\ve{q}-\mathbb{R}(\ve{t}_{\times}\ve{R}\ve{p},\theta)\ve{R}\ve{p}\|_{2}
\end{equation}
over all feature correspondences between two keyframes. 
This energy measures the Euclidean distance between $\ve{q}$ and $\ve{p}'$ where $\ve{p}'$ is the point found by rotating $\ve{p}$ with $\ve{R}$ and then with $\mathbb{R}(\ve{t}_{\times}\ve{R}\ve{p},\theta)$.
We observe that the energy in Eq.~\ref{eq:motionerror0} is hard to optimize directly because of the point-dependent angle $\theta$. 
In practice, there are about 3000 feature correspondences between two frames and each correspondence has an $\theta$, which gives a large amount of unknowns.

Instead, we directly estimate the rotation and translation between two frames by first ignoring $\theta$, and minimizing the least squares error:
\begin{equation}\label{eq:motionerror}
E_{M}(\ve{R},\ve{t})=\|\arcsin\Big({\frac{\ve{q}^{\top}\ve{t}_{\times}\ve{R}\ve{p}}{\|\ve{t}_{\times}\ve{R}\ve{p}\|}}\Big)\|_{2}
\end{equation}
over all feature correspondences.
Eq.~\ref{eq:motionerror} is a function of $\ve{R}$(three DoFs) and $\ve{t}$(three DoFs), which is much easier to solve then Eq.~\ref{eq:motionerror0}.
As shown in Fig~\ref{fig:c}, Eq.~\ref{eq:motionerror} measures the angular distance between the tracked point $\ve{q}$ and the plane that contains $\ve{R}\ve{p}$ and $\ve{t}$, which is also the angular distance between $\ve{q}$ and $\ve{p}'$.

We can then compute $\theta$ by projecting $\ve{q}$ onto the same plane with $\ve{R}\ve{p}$ and $\ve{t}$, giving the projection as $\ve{p}'$, and then measure the angular distance between $\ve{p}$ and $\ve{p}'$ as $\theta$.
We found it to be unnecessary to constrain $\ve{t}$ to be a unit vector during the optimization, as Eq.~\ref{eq:motionerror} is already normalized, and the scale of $\ve{t}$ does not affect the value of $E_{M}$. 
We therefore only normalize $\ve{t}$ after the optimization is finished.

Before minimizing Eq.~\ref{eq:motionerror}, we identify inlier feature correspondences using RANSAC with the fundamental matrix as a model, estimated by the 7-point algorithm~\cite{Hartley2004}.
The reason for not choosing the more often used 8-point algorithm is that the normalization in 8-point algorithm is inapplicable to spherical points, and the 7-point algorithm is more efficient in the presence of noise.
While the fundamental matrix is unable to give us the rotation and translation directly, it is helpful to identify feature tracks inconsistent with possible camera motion, especially for scenes with moving objects. 

We now can compute the relative motion $M_{i,{i+1}}=[\ve{R}_{i,{i+1}}, \ve{t}_{i,{i+1}}]$ between all neighboring pairs of keyframes $k_i\rightarrow k_{i+1}$ by minimizing Eq.~\ref{eq:motionerror} using a non-linear least squares solver.
We then chain rotations to align each keyframe in a global coordinate frame (e.g., that of the first keyframe):
\begin{equation}
\ve{R}_{k_i}=\prod_{j=1}^{i-1}\ve{R}_{k_j,k_{j+1}}.
\end{equation}
We use the relative translation direction $\ve{t}_{k_i,k_{i+1}}$ to compute the final warped frames and for the forward motion constraint. 



We then compute the relative camera motion for all remaining frames between a given pair of keyframes $k_i$ and $k_{i+1}$.
To do this, we set the neighboring keyframes to be the reference frame, and separately solve for the in-between frame by minimizing Eq.~(\ref{eq:motionerror}), averaging the result from both nearest keyframes (previous and next).
We evaluate the results both qualitatively and quantitatively in Sec~\ref{sec:result}.


%
\begin{figure*}
	\begin{tabular}{*{4}{c@{\hspace{3px}}}}
	\includegraphics[height=2.5cm]{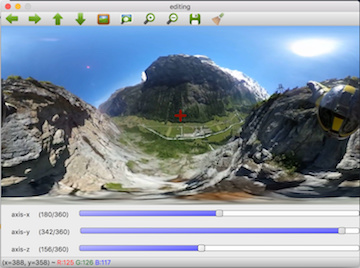} &
	\includegraphics[height=2.5cm]{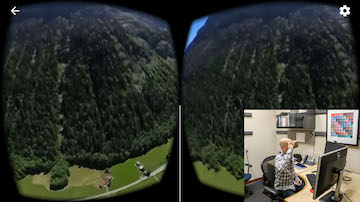} &
	\includegraphics[height=2.5cm,clip,trim= 10 0 10 0]{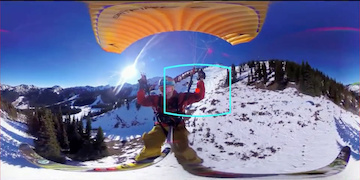} &
	\includegraphics[height=2.5cm,clip,trim=50 0 100 0]{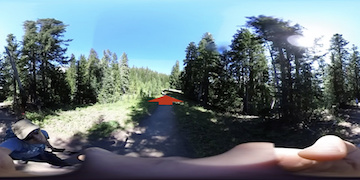} \\
	(a) & (b) & (c) & (d) \\
\end{tabular}
	\caption{\label{fig:interface} Our method can be used with a variety of input sources. Here we show some ways to provide directional constraints. A user can manually select regions on an ER projection (left), or create a guided viewing experience in a VR headset (b), or we can use automatic constraints such as saliency e.g.,~\protect\cite{supano2vid} (c), or from forward motion (Sec.~\ref{sec:motionestimation}) (d).}
\end{figure*}

\subsection{Joint Stabilization and Direction}
\label{sec:pathplanning}
Now that we have estimated the input camera path, we are able to define our joint stabilization and direction optimization, which consists of two terms:
\begin{equation}
E(W)=E_{d}(W)+E_{s}(W),
\label{eq:cinema}
\end{equation}
where $E_{d}$ is an energy that captures directional constraints and $E_{s}$ encourages the smoothness in the resulting spherical video.
$W$ is a set of camera transformations that correspond to the optimal (virtual) camera trajectory. 
In this optimization, we use a combined rotation and translation model $W$ to transform the input video, which we render using image warping.


We first solve Eq.~\ref{eq:cinema} considering $W$ restricted to \emph{rotation only} and then later solve the two terms $E_{d}(W)$ and $E_{s}(W)$ considering models that handle both rotation and translation.
In the following two sections, and implementation, we present rotation in quaternions.

\subsection{Directional Constraints}
Directional constraints can be either positive, which specify salient events that the editor would like viewers to see, or negative, indicating things that \emph{should not} be seen, for example content that is uninteresting, that contains elements that do not belong in the scene, such as the camera crew, or stitching seams.
Fig.~\ref{fig:interface} shows the various types of constraints we use. 

\paragraph{Source of Constraints}
We provide two types of user provided constraints. 
In one case, the editor simply clicks directly on the ER projection, while in the other, the editor provided a ``guided'' viewing session, where they watch the video in a VR headset, and their viewing direction is recorded over time, and used as positive constraints.
To support this, we developed an app that records the users head rotation during the playback of a video in a Google Cardboard headset.
This trajectory is then sampled at every second and used in our optimization to guide the look-at direction over the course of the video.

Our method can also be easily integrated with automatically generated constraints. 
Automatic saliency methods for 360\deg video (e.g.,~\cite{supano2vid}) determine the most visually salient regions, which can be directly interpreted as positive constraints. 
Alternately, as we are estimating 3D translation direction in Section~\ref{sec:motionestimation}, we can use this to keep the camera pointed roughly in the forward motion direction, which provides a comfortable ``first-person'' viewing experience. 
Automatic negative constraints can also be added, for example for seam locations if the camera geometry is known a priori.
We show examples of all of these types of constraints in the result section and supplemental material. 

\paragraph{Constraint Formulation}
A positive constraint is represented as a look-at point $\ve{p}_{i}$ located on a spherical video frame $f_i$.
The goal is to transform the frame $f_i$ by a rotation quaternion $\ve{q}^{W}_{f_i}$ such that $\ve{p}_i$ is as close to the true north direction as possible, thereby making it more likely to be seen.
Similarly, a negative constraint $\ve{n}_j$ located on frame $f_j$ is one that we want to avoid appearing in the front, i.e., we search for a transformation $\ve{q}^{W}_{f_j}$ to make this point appear outside of the user's likely viewing direction. 

Therefore, for a set $P$ of positive constraints, and $N$ of negative constraints, the directional term can be expressed as:
\begin{equation}
E_{d}(W)=\sum_{i \in P}\|\ve{q}^{W}_{f_i}\cdot\ve{p}_i-\ve{f}\|^{2}_{2} + \sum_{j \in N} \rho(\|\ve{q}^{W}_{f_j}\cdot\ve{n}_j-\ve{b}\|^{2}_{2}),
\label{eq:direct}
\end{equation}
where $\ve{q}^{W}_{f_i}\cdot\ve{p}_i$ denotes rotating $\ve{p}_i$ by quaternion $\ve{q}^{W}_{f_i}$, and $\ve{q}^{W}_{f_i}$ minimizes the distance between $\ve{p}_{i}$ and the front vector $\ve{f}={[0,0,1]}^{\top}$, and $\ve{q}^{W}_{f_j}$ maximizes the distance of $\ve{n}_j$ to the front vector $\ve{f}$, which is equivalent to minimizing the distance to the back vector $\ve{b}=[0,0,-1]^{\top}$.

\begin{figure}
\subfigure[L2 Loss]{\includegraphics[width=0.45\linewidth,clip,trim= 250 250 150 200]{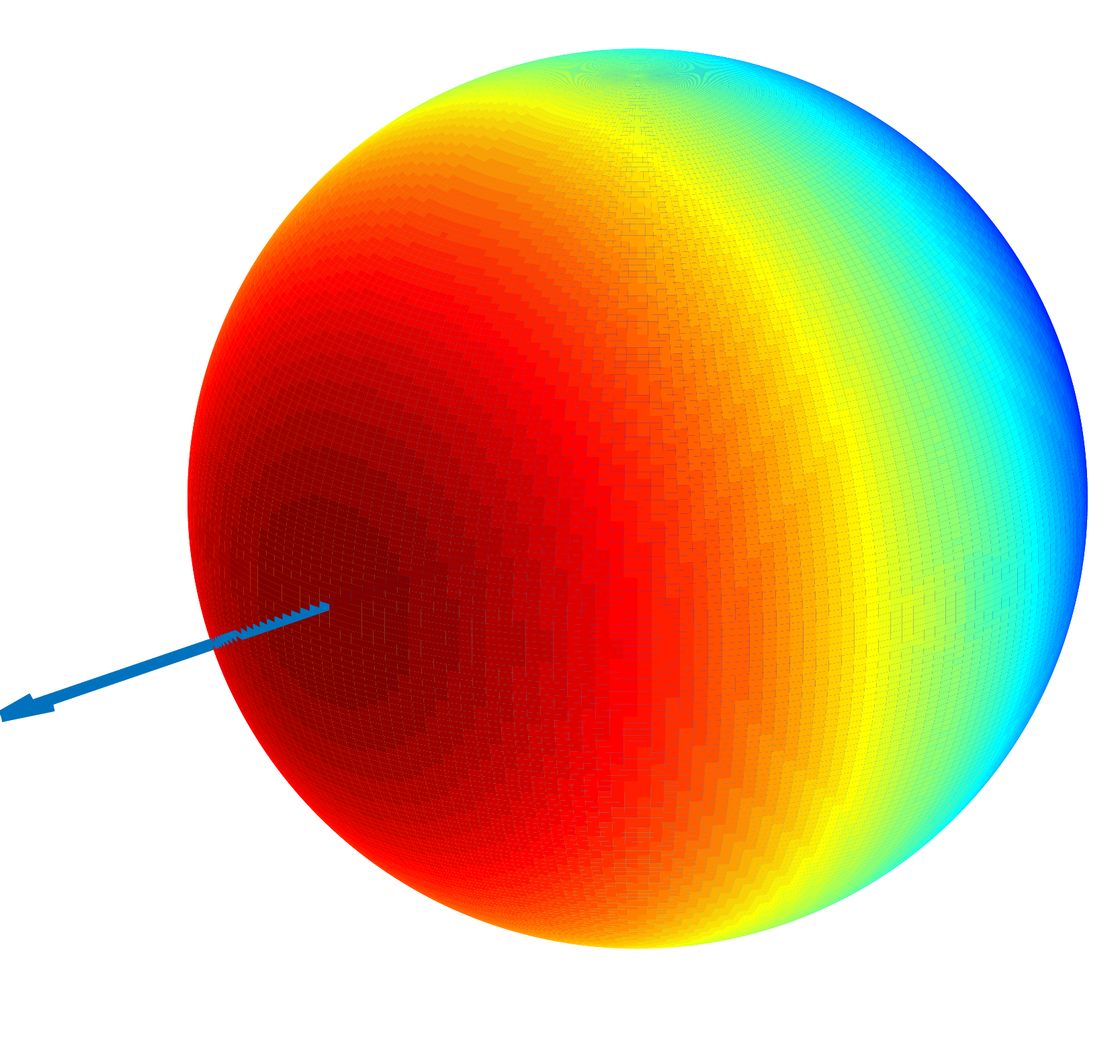}}
\subfigure[Our Loss for Negative Constraints]{\includegraphics[width=0.45\linewidth,clip,trim= 250 250 150 200]{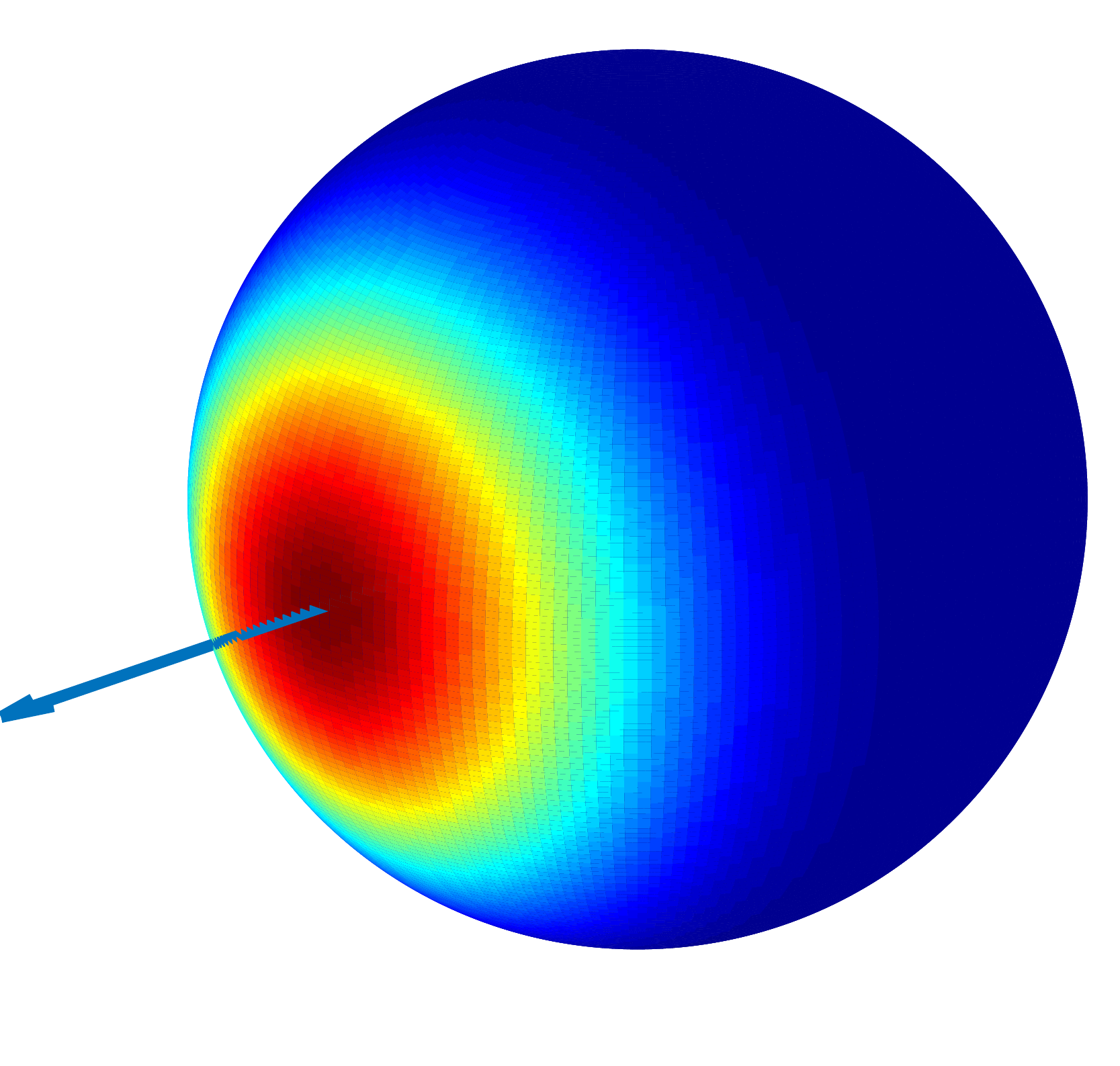}}
\caption{\tog{L2 loss vs. robust loss for negative constraints. We make the penalty close to zero as long as the negative constraint is out of visible region, and the penalty increases sharply when the negative constraint moves towards visible region. The forward direction is marked as blue arrows.}}
\label{fig:loss}
\end{figure}

Because we may only want the negative constraint to be outside the of view (rather than exactly behind the user), we use a robust loss function $\rho(x)=\alpha e^{-\frac{\beta}{x}}$ on the negative constraint.
We set $\alpha=3200$ and $\beta=26.73$, which causes $\rho(x)$ to yield a low cost until it enters a visible region for the average human field of view (which is roughly 114\degree  horizontally~\cite{strasburger2011}), after which the penalty increases sharply, as shown in Fig.~\ref{fig:loss}.

\subsection{Smoothness Constraints}
\label{sec:stability}
From Sec.~\ref{sec:motionestimation}, we have computed the global rotation $\ve{R}_{i}$ for all frames, with respect to the first frame.
\tog{
We use a smoothness model similar to ~\cite{kopf2016360}, but with small modifications. 
Importantly, we define the first-order and second-order smoothness terms directly on the estimated camera rotations instead of the feature trajectories, i.e.,
\begin{equation}\label{eq:smoothness}
E_{s}(W)={\alpha}_{1} E_{s1}(W)+{\alpha}_2E_{s2}(W),
\end{equation}
where 
\begin{equation}
E_{s1}(W)=\sum_{i=1}^{n-1} \left|\ve{q}^W_{i+1}\ve{q}_{i+1}-\ve{q}^W_i\ve{q}_{i}\right|_{p}
\end{equation} 
is the first order term and,
\begin{equation}
E_{s2}(W)={\sum_{i=1}^{n-2}}\left|(\ve{q}^W_{i+2}\ve{q}_{i+2})^{-1}\ve{q}^W_{i+1}\ve{q}_{i+1}-
									   (\ve{q}^W_{i+1}\ve{q}_{i+1})^{-1}\ve{q}^W_{i}\ve{q}_{i}\right|_{p},
\end{equation}
is the second order term. 
}

This energy is summed over all $n$ frames, and $|\ve{q}_a-\ve{q}_b|$ is the difference between two quaternion rotations, $\alpha_1$ and $\alpha_2$ are weights, and $p$ is the norm of the smoothness that we solve for.
In our implementation, we use ${\alpha}_{1}=10$ and ${\alpha}_{2}=100$ for all results shown, but these values could be changed to control the trade-off between the direction constraints and video smoothness.

\subsection{Optimization}

We can now directly minimize Eq.~\ref{eq:cinema} using the Levenberg-Marquardt algorithm in Ceres and achieve a final rotation-only directed video.
However, for efficiency, we propose the following optimization scheme which gradually propagates a good initialization, speeding up convergence.

We also observe that some axes of rotation can be confusing for the viewer.
In particular, camera movement in the roll axis is uncommon in many videos recorded on a ground plane, and can be confusing.
We therefore default to fixing the roll axis in $\psi$ and allow for rotation only in the pitch and yaw of the camera, unless otherwise specified by the user (for example in the wing-suit video (Fig.~\ref{fig:jointoptimization}), we enable rotation on the roll axis).
\togb{
We can choose between smoothness norms in our optimization, in particular prior works have used L1 and L2 norms. 
Using an L1 norm tends to give clear distinctions between fixed and moving shots, while L2 paths are smoother overall. 
We provide examples of L1 and L2 smoothed paths in the supplemental material, and allow the user to choose which norm ($p=1$ or $p=2$) they want to minimize. 
}
\subsection{Translation-Aware Transformation}
\label{sec:warping}


Until now, we have considered rotation-only transformation for $W$, which have the advantage of being quick to compute, as there are only three DoFs for each frame, and are guaranteed to not introduce any local distortion into the video.
However, as observed in prior work~\cite{kopf2016360}, camera shake often contains a significant \emph{translation} component to it, which requires local warping to compensate for parallax effects.
\tog{
One common solution is to smooth individual feature trajectories, however, this can break the spatial distance between feature trajectories on sphere and introduce geometric distortions in the output. 
Previous works have used Structure-from-Motion (SfM) or subspace constraints to generate smoothed feature trajectories while preserving scene structure, but these strategies are either computationally expensive~\cite{liu2009content} or cannot be directly applied to 360\deg videos~\cite{liu2011subspace}. 
For 360\deg videos, ~\cite{kopf2016360} represents feature points by an interpolation of six evenly distributed vertices on a sphere, and then constrains the rotation of the six vertices to be similar to avoid large distortions.
}

\tog{
Different from all these methods, we propose to solve for a virtual camera, represented by the transformation $W$ that includes rotation and translation, such that the feature trajectories as seen from this camera are smooth.  
This is possible due to the estimated per-frame 3D rotation and translation in Sec~\ref{sec:motionestimation}.
The goal of our transformation is therefore to generate smooth feature trajectories that maintain the spatial structure and directional constraints for each frame.
This approach has fewer degrees of freedom than directly optimizing for image-space warps, as it is restricted to geometrically plausible reconstructions, and we show that it can handle strong parallax better than~\cite{kopf2016360}.
At the same time, we do not require computational expensive SfM~\cite{liu2009content}.
We first describe a {\bf{spherical point projection}} model that represents transformed points as the function of our translation-aware transformation. 
We then present a modified version of the smoothness term in Eq.~\ref{eq:cinema}, that we minimize to find the transformed feature trajectories.
Finally, we can warp the image using these transformed feature trajectories in Sec.~\ref{Sec:MeshWarp}.
}

\begin{figure}
	\centering
	\includegraphics[height=0.35\linewidth,page=2]{figures/synthesize.pdf}
	\includegraphics[height=0.35\linewidth,page=1]{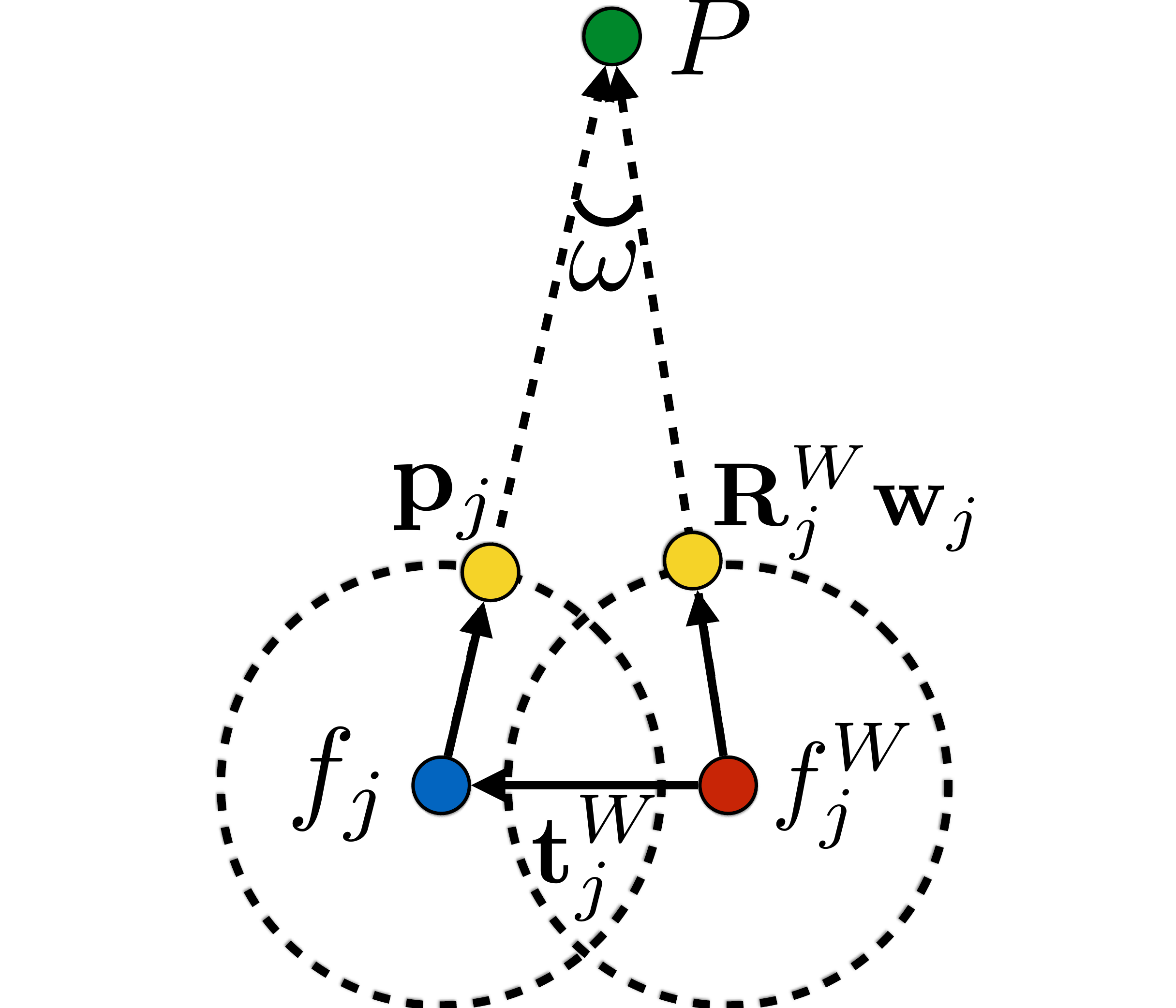}
	\caption{\tog{\label{fig:warp2} Spherical Point Projection. Left: The point projection in the input video between a frame $f_j$ and a keyframe $k_i$. Right: The point warping between an input frame $f_j$ and a target frame $f_j^{W}$.}}
\end{figure}

\paragraph{Spherical Point Projection}
\tog{
Given an inner frame $f_j$ and a neighboring keyframe $k_i$, we know the relative motion $M_{ij}=[\ve{R}_{ij},\ve{t}_{ij}]$ from Sec~\ref{sec:motionestimation}. 
As shown in Fig~\ref{fig:warp2}, we consider local coordinates centered at $f_j$, where the position of $f_j$ is the origin, the position $k_i$ is $-\ve{t}_{ij}$, and $\ve{P}$ is the 3D point corresponding to the spherical feature point $\ve{p}_j$ at a distance $d$ from $f_j$.
According to the sines rule of triangles, $d$ can be computed as:
\begin{equation}
d=\frac{\sin(\phi)}{\sin(\omega)}\|\ve{t}_{ij}\|=\frac{\sin(\phi)}{\sin(\omega)},
\label{eq:d}
\end{equation}
where $\omega$ is the angle between $\ve{p}_j$ and $\ve{R}_{ij}\ve{p}_i$ as described in Sec~\ref{sec:motionestimation},
$\phi$ is the angle between $\ve{p}_j$ and the unit translation direction $-\ve{t}_{ij}$.
Therefore, we have
\begin{equation}
\ve{P}=d\ve{p}_j=\frac{\sin(\phi)}{\sin(\omega)}\ve{p}_j.
\label{eq:P}
\end{equation}
}

\tog{
We now want to transform the point \ve{P} onto a target spherical frame $f^W_j$ by a camera transformation $W_j=[\ve{R}_{j}^W,\ve{t}_{j}^W]$. 
Since \ve{P} is already known, the transformed point $\ve{w}_j$ can be represented as a function of the transformation $W_j=[\ve{R}_{j}^W,\ve{t}_{j}^W]$ as:
\begin{equation}
\ve{w}_j={(\ve{R}_{j}^{W})}^{\top}(\frac{\ve{P}-\ve{t}_{j}^W}{\|\ve{P}-\ve{t}_{j}^W\|}).
\label{eq:ps}
\end{equation}
}
\tog{Dividing $\sin(\omega)$ in Eq.~\ref{eq:d} and Eq.~\ref{eq:P} makes $\ve{P}$ numerically unstable when $\sin(\omega)$ is close or equal to zero. 
This issue happens when the translation is near zero, or $\ve{P}$ is a far scene point. 
To avoid this problem, we scale both $\ve{P}$ and $\ve{t}_{j}^{W}$ in Eq.~\ref{eq:ps}, by $\sin(\omega)$, which leads to the same result but one that is numerically more stable.}

\paragraph{Translation-Aware Optimization}
\tog{We next introduce how to optimize for the final smoothed and directed camera transformations. 
By representing a transformed point $\ve{w}_j$ as the function of a rotation $\ve{R}_{j}^{W}$ and a translation $\ve{t}_{j}^W$ in Eq.~\ref{eq:ps2}, we can extend the rotation only transformation in Eq.~\ref{eq:cinema} to a full rotation and translation transformation:
	\begin{equation}
	\resizebox{\linewidth}{!}{
		$\ve{w}_j={(\ve{R}_{j}^{W})}^{\top}(\frac{\sin(\omega)(\ve{P}-\ve{t}_{j}^W)}{\|\sin(\omega)(\ve{P}-\ve{t}_{j}^W)\|})={(\ve{R}_{j}^{W})}^{\top}(\frac{\sin(\phi)\ve{p}_j-\sin(\omega)\ve{t}_{j}^W}{\|\sin(\phi)\ve{p}_j-\sin(\omega)\ve{t}_{j}^W\|})$.}
	\label{eq:ps2}
	\end{equation}
The optimization for the joint rotation and translation $W$:
\begin{equation}
E_s(W)=\sum^{|T|}_{i=1}\Big(\alpha_1\sum^{e_i-1}_{j=s_i}\|\ve{w}^{i}_{j}-\ve{w}^{i}_{j+1}\|+\alpha_2\sum^{e_i-2}_{j=s_i}\|\ve{w}^{i}_{j+2}-2\ve{w}^{i}_{j+1}+\ve{w}^{i}_{j}\|\Big)
\label{eq:smoothft}
\end{equation} 
which is an objective function over the unknown camera transformation, $s_i$ and $e_i$ are the starting and ending frame of the $i$-th feature trajectory as in Eq.~\ref{eq:ft}, $\alpha_1$ and $\alpha_2$ are the same weight between first-order and second-order smoothness as in Eq.~\ref{eq:smoothness}. 
}

\begin{figure}
	\centering
	\includegraphics[width=\linewidth,page=1]{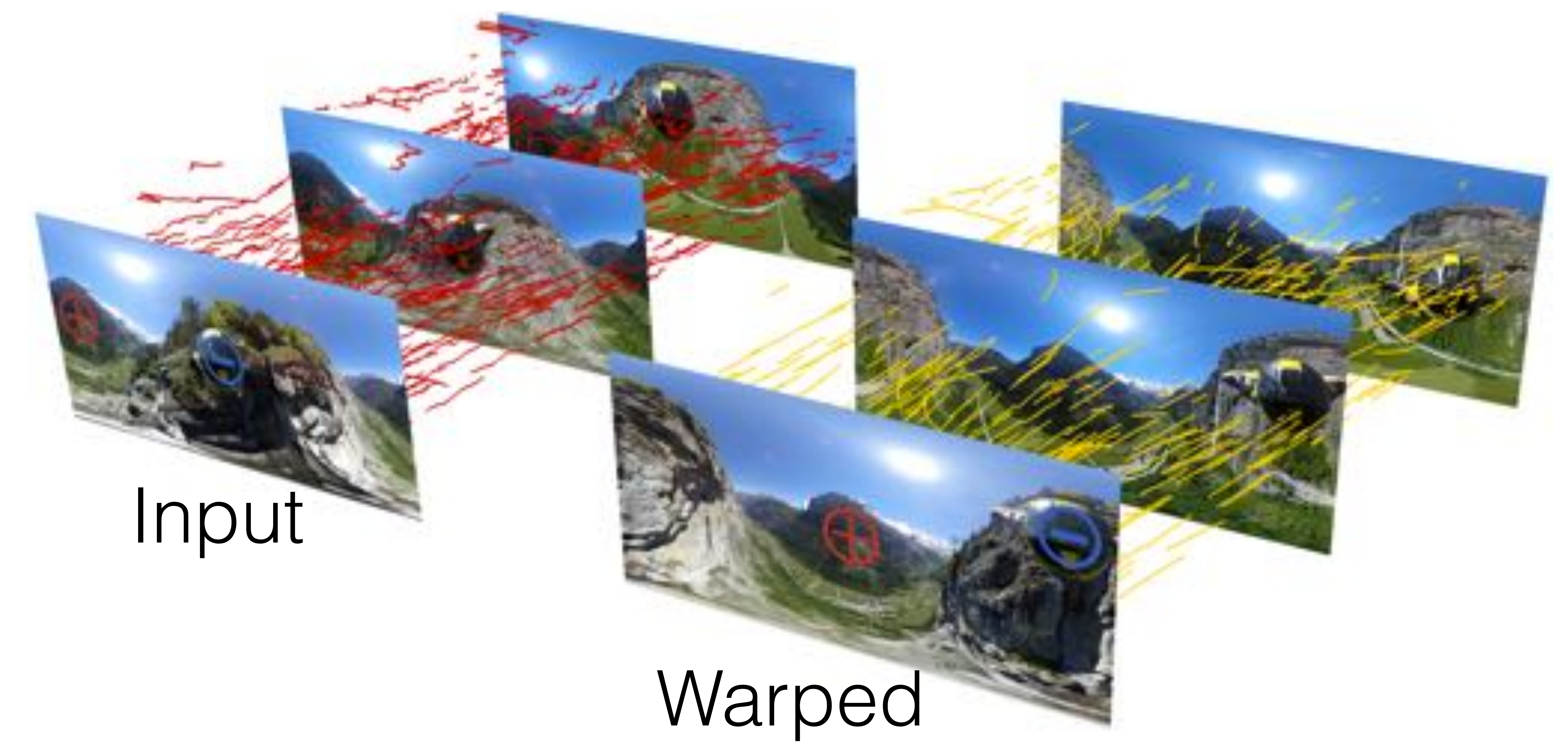}
	\caption{\label{fig:3dwarp} \tog{Translation-Aware Optimization. The input trajectories (red) are shaky while the optimized trajectories (yellow) are smoother. At the same time, the positive constraint (red $'+'$ circle) is transformed to the front, while the negative constraint (blue $'-'$ circle) is transformed to the back of the sphere.}}
\end{figure}

\tog{
The final objective function combines the smoothness term Eq.~\ref{eq:smoothft} and the directional term, (Eq.~\ref{eq:cinema}).
And minimizing this yields the transformation $W$ that represents the collection of rotation and translation $W={\{[\ve{R}_{i}^{W},\ve{t}_{i}^{W}]\}}_{i=1\cdots n}$ that transform all the $n$ frames to a smoother and better directed camera path.
After $W$ is estimated, we transform the points of feature trajectories by Eq.~\ref{eq:ps2}.
In the example illustrated in Fig.~\ref{fig:3dwarp}, we can see that the transformed feature trajectories are much smoother than the original input, while satisfying directional constraints. 
}


\begin{figure}
	\centering
	\includegraphics[width=0.8\linewidth,trim= 0 20 0 20]{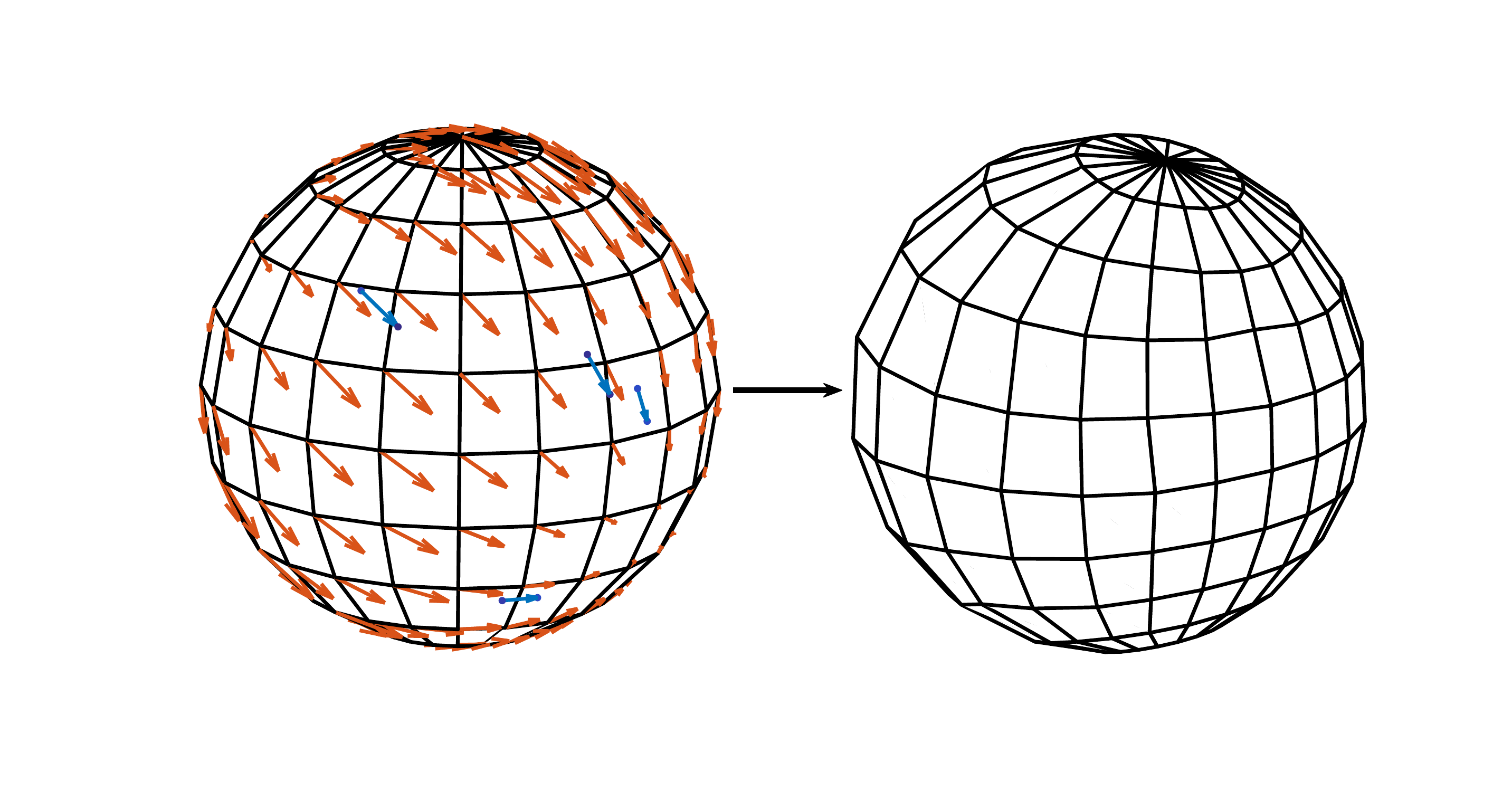}
	\caption{\label{fig:warp}Spherical Mesh Warping. A set of feature correspondences (blue arrows) are used to drive a warp computed over the vertices of a sphere. The computed offset vectors at each vertex (red arrows) are used to warp the vertices to the sphere on the right.}
\end{figure}

\subsection{3D Spherical Mesh Warping}
\label{Sec:MeshWarp}
We now render these transformations by spherical mesh warping.
For each frame $f_j$ we have $\ve{p}$ as the original feature point in the input and $\ve{w}$ as the corresponding points in the transformed frame $f_j^W$.
We ignore the superscript $W$ and subscript $j$ since we consider warping now only for single frames.
Based on the point correspondences $\ve{p}\rightarrow\ve{w}$, we can re-render an input image as if the image was captured by a camera physically located on a well directed and smooth 3D camera path without requiring full 3D reconstruction, which can be slow and unreliable~\cite{liu2009content}. 
As shown in Fig~\ref{fig:warp}, the general idea of our spherical mesh warping is to propagate the point-wise correspondence to each mesh vertex in the image and then warp the spherical image based on these vertices.
Due to our 3D structure preservation, we can allow for a higher degree of deformation than prior work (we use a $20\times10$ mesh), while avoiding geometric distortion. 
Given the high mesh resolution, we found it to be sufficient to use a regular grid defined on an equirectangular map, however different spherical tessellations could be used if desired.

As described in Sec.~\ref{sec:motionestimation}, the translation induced motion can be represented by feature dependent axis and angle. 
Since we've already estimated the transformation rotation $\ve{R}$ and translation $\ve{t}$, we know the axis as $\ve{t}_{\times}\ve{R}\ve{p}$ for each feature point, and we get the angle $\omega$ in the same way as in Sec.~\ref{sec:motionestimation}.

We now have the rotation and translation between the input and warped image and have estimated the angle $\omega$ (which is a parametrization of the point depth) for each feature in the frame. 
To propagate the warping motion from a set of feature points $F$ to a set of \emph{mesh vertices} $V$ in the corresponding spherical mesh warp, we apply the rotation $\ve{R}$ to all vertices and then solve for $E_\omega$, the field of translation angles for all vertices by a 1D minimization parameterized \emph{only} on the angle $\omega$.
\begin{equation}
E_\omega=\sum_{p\in F}\|{\omega}_{p}-\ve{b}_{p}^{\top}\ve{v}_{p}\|^{2}_{2}+\sum_{i,j\in V}\|{v}_{i}-{v}_{j}\|^{2}_{2},
\label{eq:1doptimize}
\end{equation}
where $p$ belongs to the set of feature points, $i$ and $j \in V$ are neighboring vertices on sphere, $\ve{v}_{p}$ are the angles for the four vertices that cover the $p$-th feature point~\cite{liu2009content}, 
and $\ve{b}_{p}^{\top}$ is the corresponding bilinear interpolation coefficients for the four vertices.
The spatial smoothness term $\sum\|{v}_{i}-{v}_{j}\|^{2}_{2}$ guarantees that angles are spatially smooth and avoids local distortion. 

\begin{figure}
	\centering
	\includegraphics[height=0.23\linewidth]{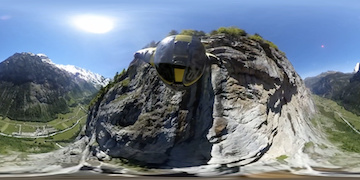}
	\includegraphics[height=0.23\linewidth]{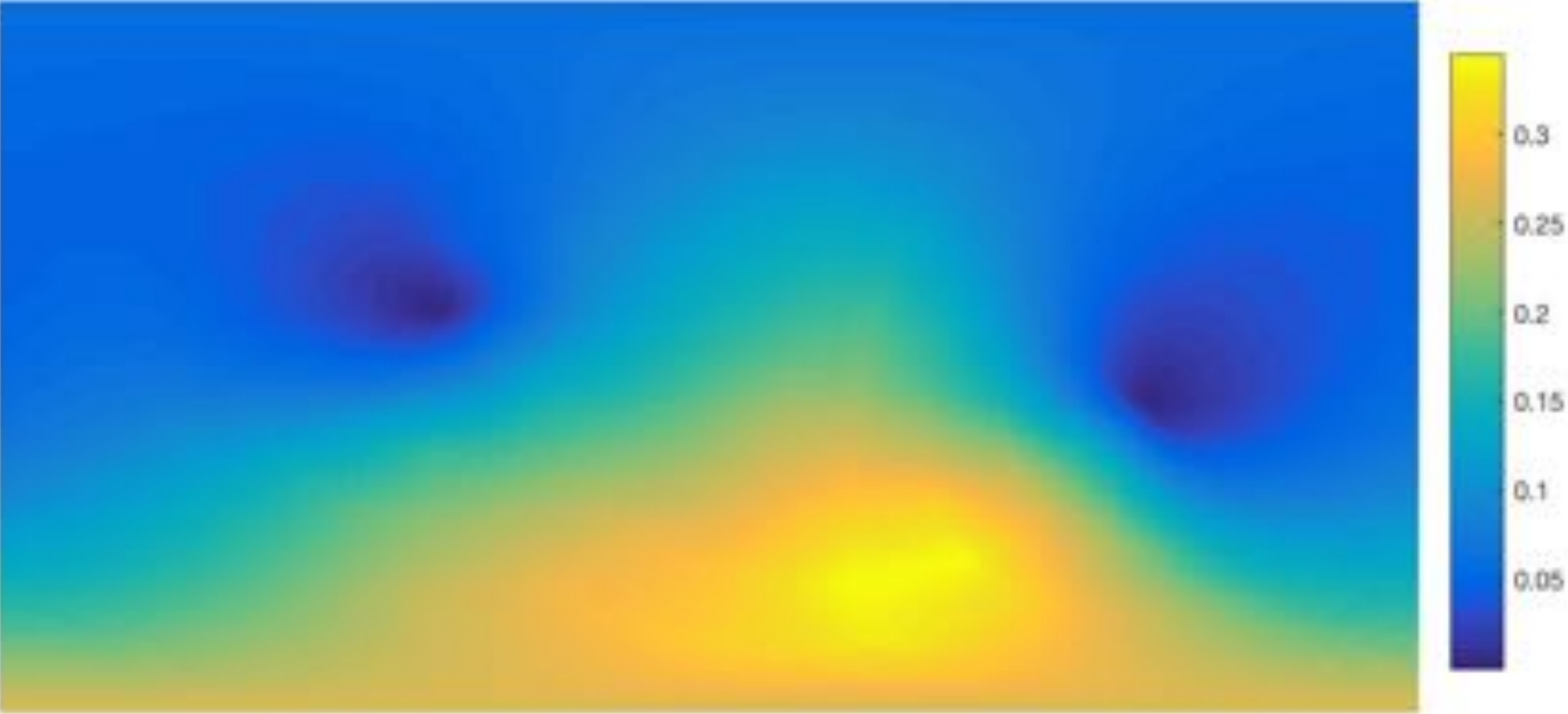}
	\caption{\label{fig:angle}A warped image and the interpolated angle $\phi$ shown for all pixels. This demonstrates how the warping angles have structure roughly similar to the inverse depth map but smooth, e.g., pixels closer to the camera on the mountain have larger value than farther pixels under the same camera motion.}
\end{figure}

We solve Eq.~\ref{eq:1doptimize} in a least squares sense, giving us the final output position for each vertex and use these vertex positions to render a warped sphere (Fig.~\ref{fig:warp}).
The 1D optimization defined in Eq.~\ref{eq:1doptimize} has the advantage of being efficient to compute, while restricting our solution to warps that are geometrically regularized.
\tog{
We bi-linearly interpolate the angle to all pixels and visualize it in Fig~\ref{fig:angle}.
Because the warping must only correct for residual motion, we observed, similar to prior work~\cite{kopf2016360}, that the warping function is largely smooth, and as such we have found this approach to be robust to input videos and parameter settings. 
}



\subsection{Implementation}


Our method was implemented in C++. 
We solve Eq.~\ref{eq:motionerror} and Eq.~\ref{eq:cinema} using the Levenberg-Marquardt algorithm in Ceres. 
We fixed parameters for all the results shown here, although if desired we can easily control the amount of smoothing by changing $\alpha_1$ and $\alpha_2$.
We report average running times for the different parts of our method in Tab.~\ref{tab:runningtime} computed on a 2015 Macbook Pro laptop with 2.5GHz i7 CPU and 16GB memory, on HD (1920 $\times$ 960) equirectangular video frames.

\begin{table}[h]
	\centering
	\definecolor{rowblue}{RGB}{220,230,240}
	\setlength{\tabcolsep}{5pt} 
	\setlength\extrarowheight{1.8pt}
	\rowcolors{2}{rowblue}{white}
	\resizebox{.95\linewidth}{!}{	
		\begin{tabular}{lr}
			\toprule
			Section & \parbox{3cm}{\begin{flushright}Average running time in ms/frame\end{flushright}} \\
			\midrule
			\midrule
			Ingest and cube map conversion & 5 \\ 
			Feature tracking and keyframe selection (Sec.~\ref{sec:feature}) & 23 \\
			Relative motion estimation (Eq.~\ref{eq:motionerror}) & 10 \\
			Re-cinematography computation (Eq.~\ref{eq:cinema}) & 1 \\ 
			Full deformable warping (Eq.~\ref{eq:1doptimize}) & 10 \\
			Rendering in OpenGL & 6 \\
			\midrule
			Total & 55 \\
			\bottomrule
		\end{tabular}
	}
	\vspace{.3cm}
	\normalsize
	\caption{\label{tab:runningtime} Running times for different parts of our method, computed on 1920 $\times$ 960 video frames.}
\end{table}

\begin{figure}[t]
	\centering
	\begin{tabular}{*{4}{c@{\hspace{3px}}}}
		\begin{sideways}\hspace{3.0em}Tilt\end{sideways}& 
		\includegraphics[height=2.3cm]{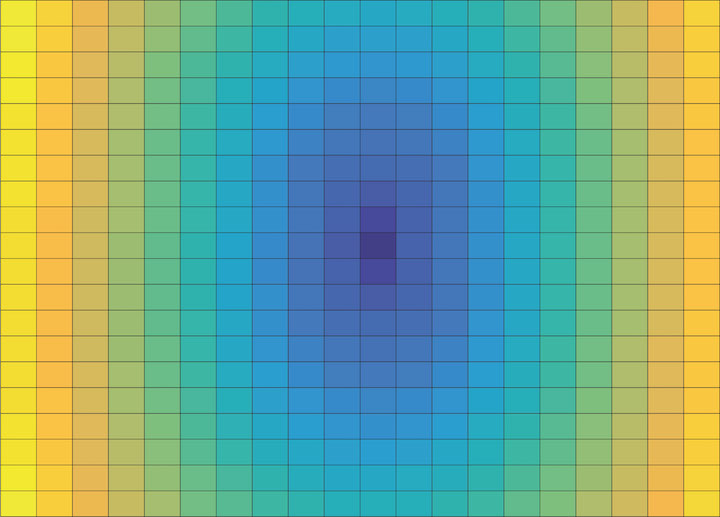} &
		\begin{sideways}\hspace{3.5em}Z\end{sideways}& 
		\includegraphics[height=2.3cm]{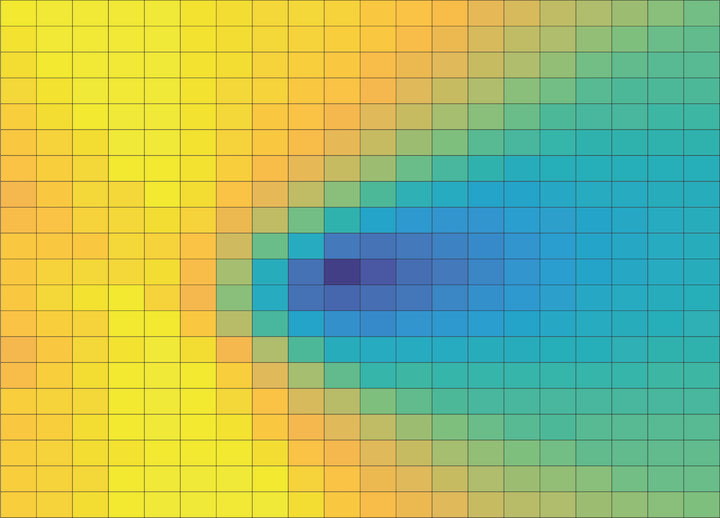} \\
		& Pan & & X\\
	\end{tabular}
	\caption{\label{fig:error}Representative slices of the energy surface formed by Eq.~\ref{eq:motionerror}, with blue showing areas of lower energy. Here we show 2D slices of the original 6D error space. This visualization indicates that the surface is largely smooth, which yields a robust and efficient optimization, despite the nonlinear energy terms.}
\end{figure}

\section{Results and Evaluation}
\label{sec:result}

\subsection{Motion Estimation Evaluation}
We first compare our motion estimation to the commonly used 5-point algorithm~\cite{5PT}.
As opposed to the 5-point algorithm which estimates an essential matrix, we directly solve for $\ve{R}$ and $\ve{t}$ using non-linear least squares.
We solve this minimization using Levenberg-Marquardt optimization~\cite{ceres-solver} \tog{with manually derived analytic derivatives which we show in the Appendix A.} 
We empirically found this to converge by 10 iterations, taking roughly 20ms to estimate the motion for a pair of cameras with about 3000 feature correspondences.
Surprisingly, the optimization converges to a reasonable solution even when initialized with an identity rotation matrix and a random unit translation vector. 
To understand why, we visualize the error space by uniform sampling (two slices of which are shown in Figure~\ref{fig:error}), which shows that the energy function is close to convex, making it efficient to solve robustly.
We also found that our motion estimation works well even in cases with pure rotation, which is most likely due to the fact that our method uses all feature points available in a 360\deg image, which helps to disambiguate rotation and translation motion.
Compared to the 5-point algorithm~\cite{5PT}, where epipolar constraint are enforced exactly (Eq.~\ref{eq:rotation} equals to zero) within each 5-point group and the solution is selected as the one with maximum inliers, our method seeks for a solution with a minimum energy over all points, which benefits from a greater number of inliers, improving the accuracy of the motion estimation and mesh-based image warping.


\begin{figure}[t]
		\definecolor{myblue}{rgb}{.1,0.3,0.9}
\definecolor{myred}{rgb}{.9,0.3,0.1}
\definecolor{mygreen}{rgb}{.1,0.6,0.1}
	\begin{tikzpicture}
	
	\pgfplotsset{
		scale only axis,
		width=7cm,
		height=3.5cm,
		legend style={at={(0.85,0.85)},anchor=north}
	}
	
	\begin{axis}[
	xlabel=Noise Added (deg),
	ylabel= Rotation Error (deg),
	grid=both,
	grid style={line width=.1pt, draw=gray!20},
	]
	\addplot[line width=.5mm,smooth,mark=*,myred]
	coordinates{
 (0,0)
 (0.1432,0.0063)
 (0.2865,0.0127)
 (0.4297,0.0196)
 (0.5730,0.0247)
 (0.7162,0.0358)
 (0.8594,0.0459)
	}; \label{Ours}
	\addlegendentry{Ours}

	\addplot[line width=.5mm,smooth,mark=*,mygreen]
	coordinates{
		(0,1.1256)
		(0.1432,1.1291)
		(0.2865,1.1371)
		(0.4297,1.1194)
		(0.5730,1.1715)
		(0.7162,1.1576)
		(0.8594,1.1478)
	};  \label{Rotation}
	\addlegendentry{$\ve{R}$ only}
	
	\addplot[line width=.5mm,smooth,mark=*,myblue]
	coordinates{
(0,0)
(0.1432,0.1549)
(0.2865,0.2443)
(0.4297,0.3121)
(0.5730,0.3439)
(0.7162,0.3807)
(0.8594,0.4080)
	};  \label{5PT}
	\addlegendentry{5-point}

	\end{axis}
	\end{tikzpicture}
	
		\begin{tikzpicture}
	
	\pgfplotsset{
		scale only axis,
		width=7cm,
		height=2.5cm,
		legend style={at={(0.85,0.70)},anchor=north}
	}
	
	\begin{axis}[
	xlabel=Noise Added (deg),
	ylabel= Translation Error (deg),
	grid=both,
	grid style={line width=.1pt, draw=gray!20},
	]
	\addplot[line width=.5mm,smooth,mark=*,myred]
	coordinates{
		(0,0.0164)
		(0.1432,0.0215)
		(0.2865,0.0379)
		(0.4297,0.0531)
		(0.5730,0.0723)
		(0.7162,0.1007)
		(0.8594,0.1350)
	}; \label{Ours}
	\addlegendentry{Ours}	
	
	\addplot[line width=.5mm,smooth,mark=*,myblue]
	coordinates{
		(0,0)
		(0.1432,0.4480)
		(0.2865,0.6854)
		(0.4297,0.8790)
		(0.5730,0.9912)
		(0.7162,1.2151)
		(0.8594,1.2060)
	};  \label{5PT}
	\addlegendentry{5-point}

	\end{axis}
	\end{tikzpicture}

	\begin{tikzpicture}
	\pgfplotsset{
		scale only axis,
		width=7cm,
		height=3.5cm,
		legend style={at={(0.85,0.85)},anchor=north}
	}
	
	


	\begin{axis}[
	xlabel=FoV(deg),
	ylabel= Rotation Error (deg),
	grid=both,
	grid style={line width=.1pt, draw=gray!20},
	]
	\addplot[line width=.5mm,smooth,mark=*,myred]
	coordinates{
		(60,0.758038)
		(120,0.0910256)
		(180,0.014939)
		(360,0.0129926)
	}; \label{Ours}
	\addlegendentry{Ours}	
	
	\addplot[line width=.5mm,smooth,mark=*,myblue]
	coordinates{
		(60,1.55689)
		(120,0.789705)
		(180,0.55792)
		(360,0.401221)
	};  \label{5PT}
	\addlegendentry{5-point}

	\end{axis}
	\end{tikzpicture}
	\caption{\label{fig:5ptquan}\tog{Validation of our proposed method on synthetic data. Top: A rotation-only model $\ve{R}$ (green) performs significantly worse than our full approach that models both $\ve{R}$ and 3D translation direction $\ve{t}$ (red), which outperforms the commonly-used 5-point algorithm (blue) in the presence of increasing noise. Middle: We observe a similar difference when comparing our recovered 3D translation direction to the 5-point algorithm. Bottom: When comparing the result quality at different FoV settings, we can see that our direct estimation has consistently less error.}}
\end{figure}
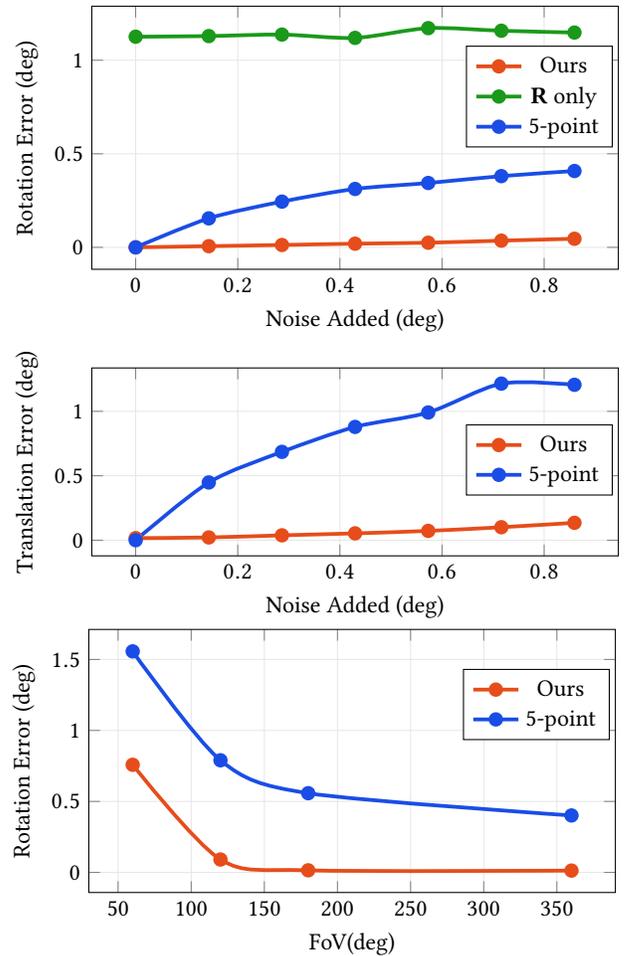

\begin{figure}[t]
	\advance\leftskip-3px
	\hspace{-10px}
	\begin{tabular}{*{3}{c@{\hspace{3px}}}}	
		\includegraphics[clip,width=0.33\linewidth]{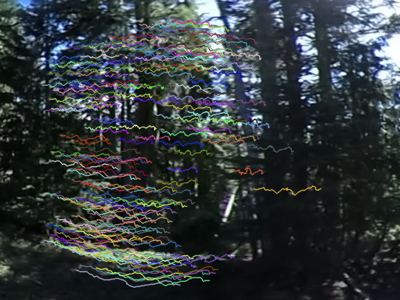} & 
		\includegraphics[clip,width=0.33\linewidth]{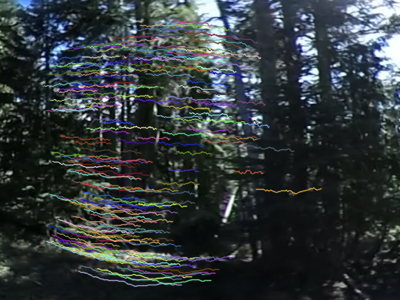} & 
		\includegraphics[clip,width=0.33\linewidth]{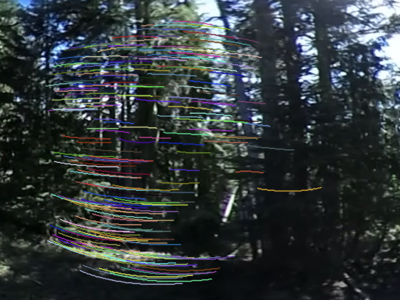}\\ 
		rotation only &5-point & Ours \\ 
	\end{tabular}
	\caption{\label{fig:5ptqual}Visualization of feature tracks after running our full method using the 5-point algorithm for estimating 3D rotation and translation (left) vs using our proposed approach (right), showing that our result is smoother. Please see the supplemental video for a comparison.}
\end{figure}

We next evaluate the motion estimation quality in the presence of noise in the feature correspondences.
A qualitative visualization of feature trajectories after stabilization using our method in place of the 5-point algorithm is shown in Fig.~\ref{fig:5ptqual}, with additional examples in Sec.~\ref{sec:result}. 
Quantitative evaluation on real world data is challenging due to the difficulty of collecting ground truth data for 3D pose estimation.
We therefore employ the same validation technique used by prior motion estimation works~\cite{5PT,Kneip2012,Kneip2013}.
The approach is to use synthetic data, where the first camera is fixed at the origin with identity rotation and a second camera chosen at a random position at most $\tau$ units from the first with a relative rotation generated from random Euler angles bounded to $\kappa$\deg.
We then create a uniformly distributed 3D point cloud with fixed maximum distance to the origin $\gamma$.
Feature correspondences can then be computed by projecting the 3D points into the two spherical cameras with added noise.
Finally, the relative motion of the cameras are computed from this data and compared to the known relative camera motion.

For our experiments we use $\tau=2$, $\kappa=30$\deg, $\gamma=8$.
We fix the outlier percentage to 10\% and solve for the relative poses 1000 times at each noise level, recording the final average accuracy. 
\tog{
	As shown in Fig~\ref{fig:5ptquan} our direct motion estimation is more accurate than the 5-point algorithm for 360\deg video, and is more robust to increasing noise levels due to the fact that it can consider a larger number of background tracks. 
	Estimating both rotation and translation also enables us to derive the 3D spherical warping proposed in Sec.~\ref{sec:warping}.
We also found the approach to be particularly robust to the field of view (FoV) parameter for traditional video, which allowed us to use a rough estimate of 120\deg horizontal FoV for all wide-angle videos, with the vertical FoV dependent on the aspect ratio.}

\begin{figure}
	\subfigure[First-order Smoothness]{\includegraphics[width=0.495\linewidth]{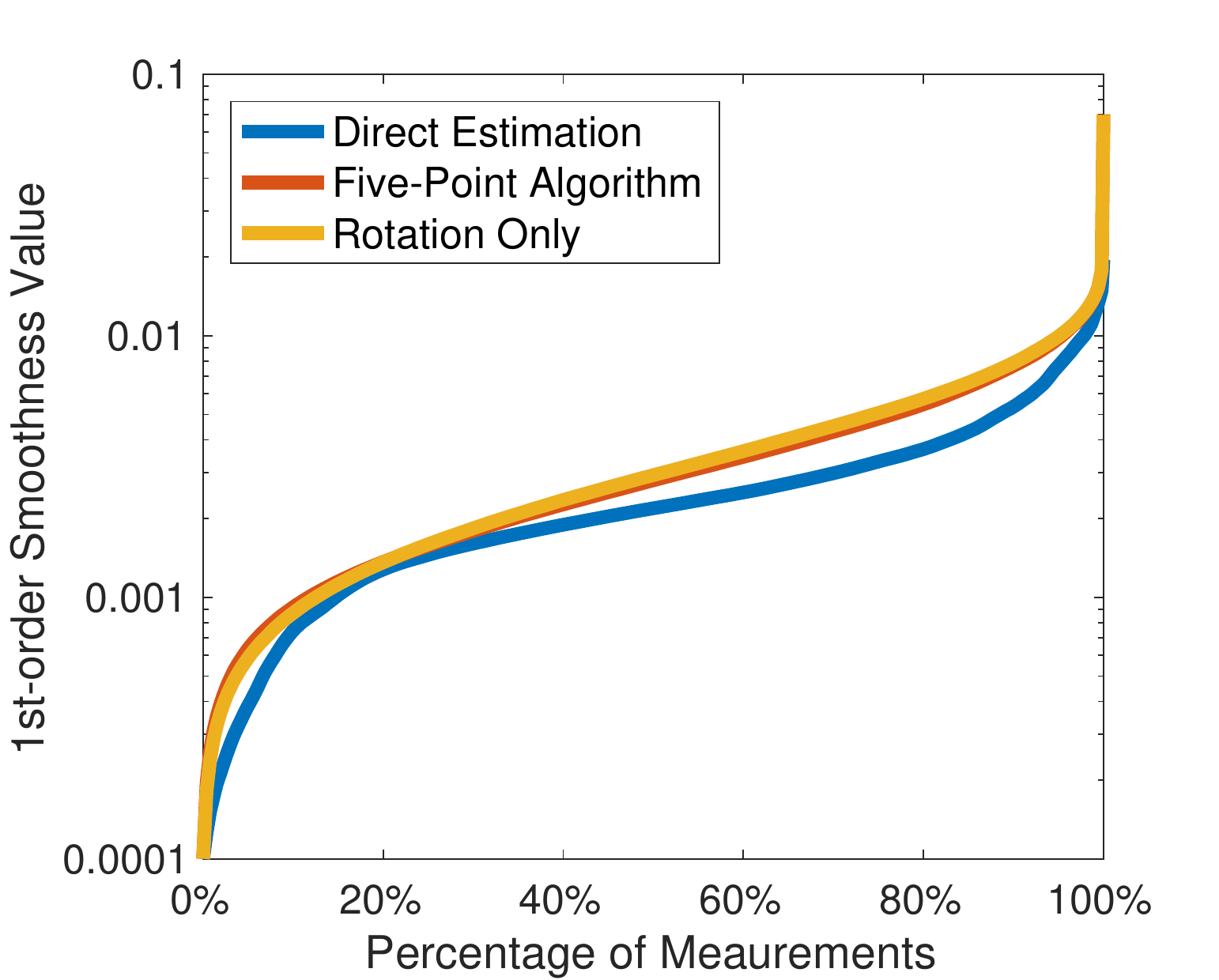}}
	\subfigure[Second-order Smoothness]{\includegraphics[width=0.495\linewidth]{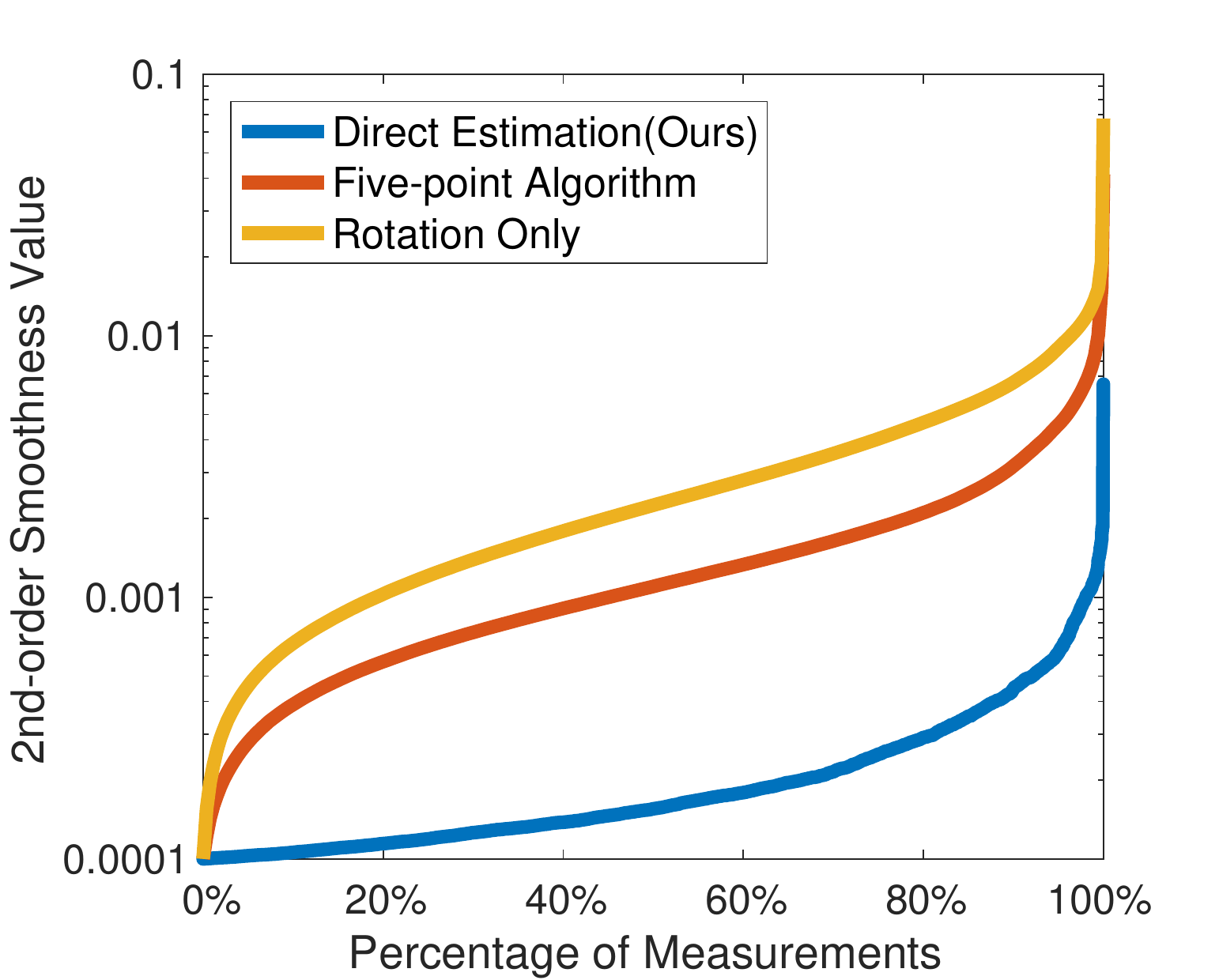}}
	\caption{\tog{Quantitative comparison of first-order and second-order smoothness by different motion estimation methods. The curves show the cumulative distribution function of (a) first-order and (b) second-order smoothness costs for the feature trajectories stabilized by different motion estimation algorithms. The y-axis represents the value of the first and second-order smoothness, and x-axis represents the percentage of the values larger than a corresponding y value.}}
	\label{fig:mecompare}
\end{figure}
\tog{In addition to the quantitative evaluation of our motion estimation in Sec.~\ref{sec:motionestimation}, we also show the smoothness of final results achieved by different motion estimation strategies.
Similar to~\cite{kopf2016360}, we collect all feature trajectories and evaluate the value of first-order and second-order smoothness terms in Eq.~\ref{eq:smoothft}.
Fig.~\ref{fig:mecompare} shows the cumulative distribution functions (CDFs) for these quantities by different motion estimation methods. 
We can see that feature trajectories optimized by our translation-aware transformation are smoother than 5-point algorithm, and that, as also noted in~\cite{kopf2016360} that rotation only model can not stabilize the video sufficiently.}


\begin{figure*}[t]
	\newcommand{\thumbnailheight}{1.7cm}
	\subfigure[BACKYARD]{\includegraphics[clip,trim=18.75 0 18.75 0, height=\thumbnailheight]{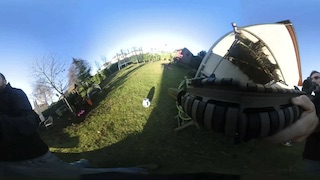}}
	\subfigure[BIKING]{\includegraphics[clip,trim=18.75 0 18.75 0,height=\thumbnailheight]{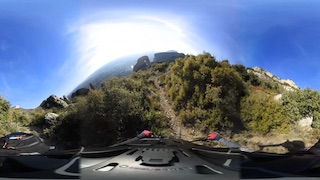}}
	\subfigure[WINGSUIT]{\includegraphics[clip,trim=18.75 0 18.75 0,height=\thumbnailheight]{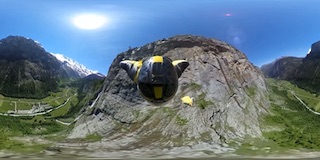}}
	\subfigure[MARKET]{\includegraphics[clip,trim=18.75 0 18.75 0,height=\thumbnailheight]{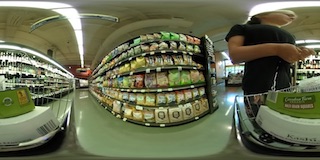}}
	\subfigure[GUNDAM]{\includegraphics[clip,trim=18.75 0 18.75 0,height=\thumbnailheight]{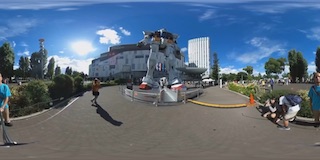}}
	\subfigure[HIKING]{\includegraphics[clip,trim=18.75 0 18.75 0,height=\thumbnailheight]{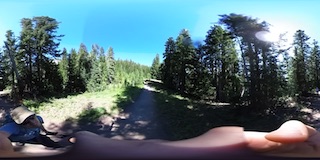}}
	\subfigure[JUNGLE]{\includegraphics[clip,trim=18.75 0 18.75 0,height=\thumbnailheight]{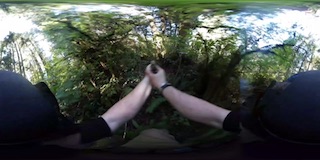}}
	\subfigure[KENDO]{\includegraphics[clip,trim=18.75 0 18.75 0,height=\thumbnailheight]{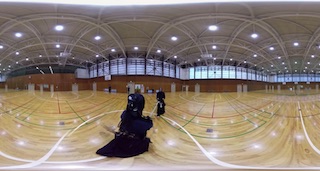}}
	\subfigure[SLIDING]{\includegraphics[clip,trim=18.75 0 18.75 0,height=\thumbnailheight]{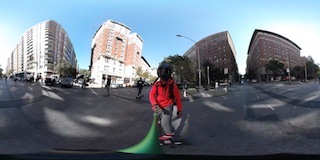}}
	\subfigure[HIKING2]{\includegraphics[clip,trim=18.75 0 18.75 0,height=\thumbnailheight]{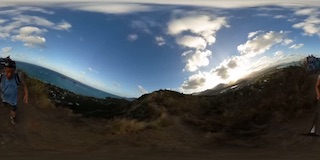}}
	\subfigure[HIKING3]{\includegraphics[clip,trim=18.75 0 18.75 0,height=\thumbnailheight]{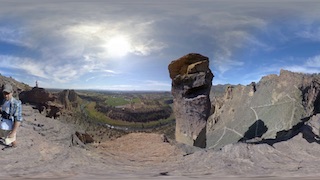}}
	\subfigure[SOCCER]{\includegraphics[clip,trim=18.75 0 18.75 0,height=\thumbnailheight]{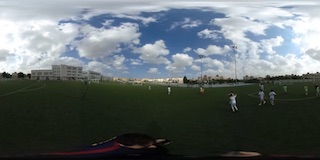}}
	\subfigure[CRANE$^*$]{\includegraphics[clip,trim=18.75 0 18.75 0,height=\thumbnailheight]{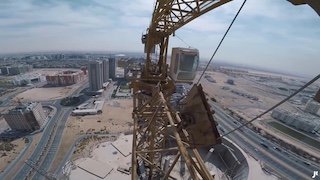}}
	\subfigure[HAWK$^*$]{\includegraphics[clip,trim=18.75 0 18.75 0,height=\thumbnailheight]{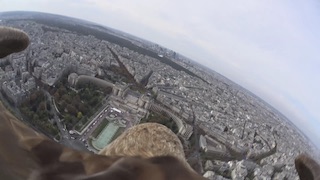}}
	\subfigure[LEGO$^*$]{\includegraphics[clip,trim=18.75 0 18.75 0,height=\thumbnailheight]{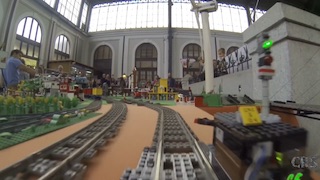}}
	\subfigure[PARKOUR$^*$]{\includegraphics[clip,trim=18.75 0 18.75 0,height=\thumbnailheight]{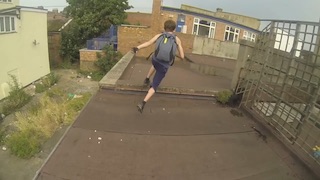}}
	\subfigure[SPEEDFLYING$^*$]{\includegraphics[clip,trim=18.75 0 18.75 0,height=\thumbnailheight]{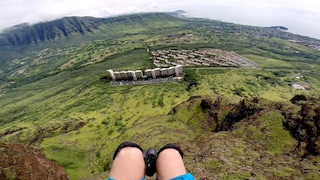}}
	\subfigure[WOLF$^*$]{\includegraphics[clip,trim=18.75 0 18.75 0,height=\thumbnailheight]{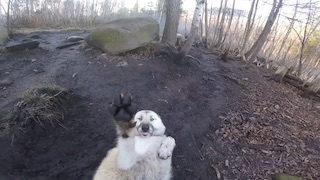}}
	\caption{\tog{360\deg Video Sequences. Thumbnails from the video sequences used for results. These include 12 360 video sequences from Youtube, and the sequences HIKING2, HIKING3 and SOCCER from the Pano2Vid dataset. Additional wide-angle videos are delineated with an $^*$.}}

\end{figure*}

\subsection{Qualitative Evaluation}
\begin{figure}
\subfigure[L1 smoothness]{\includegraphics[width=0.32\linewidth,page=1]{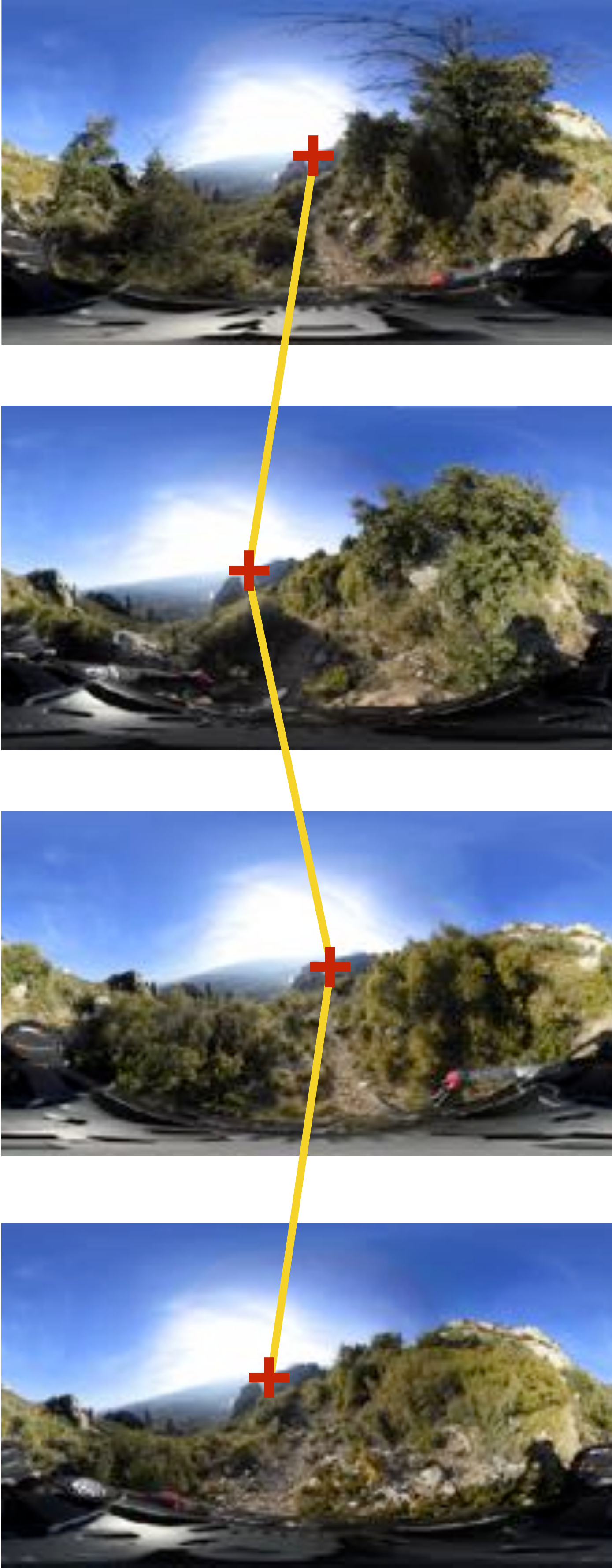}}
\subfigure[L1 smoothness with third order term]{\includegraphics[width=0.32\linewidth,page=2]{figures/bike_compare.pdf}}
\subfigure[L2 smoothness]{\includegraphics[width=0.32\linewidth,page=3]{figures/bike_compare.pdf}}
\caption{\tog{We show the horizontal movement of a fixed scene point with different smoothness constraints. We can see that in this case the L2 smoothness term generates a more stable result than L1 smoothness.}}
\label{fig:bike_l1vl2}
\end{figure}

\tog{In this section, we perform ablation studies and compare alternatives for each component of our method.
For evaluation, we collected 12 360\deg and 8 wide angle videos from YouTube, as well as 3 360\deg videos from the Pano2Vid dataset~\cite{supano2vid}.
The resolution ranges from $1920\times960$ to $3840\times2160$, for 360\deg videos, and $1280\times720$ to $3840\times2160$ for wide angle videos, with unknown FoVs. 
We then trim long videos such that the final duration ranges from 30-60 seconds, which is long enough for viewers to fully explore the video and much longer than prior datasets used for narrow FOV video stabilization~\cite{liu2013bundled}.}
In Fig.\ref{fig:bike_l1vl2}, we compare smoothness norms by visualizing the path of a scene point across frames.
Please see the supplemental video for reference.

\tog{\review{We present all video results in the supplemental material in equirectangular format, which we encourage to be viewed on a VR headset if possible.
If this is not possible, we provide a link to a desktop viewer where the videos can be explored by mouse as well as a normal FoV video that presents select results.}}

\begin{figure}[t]
	\begin{tabular}{*{2}{c@{\hspace{3px}}}}
		\begin{sideways}\hspace{2.8em}Input\end{sideways}& 
		\includegraphics[width=\linewidth,page=1]{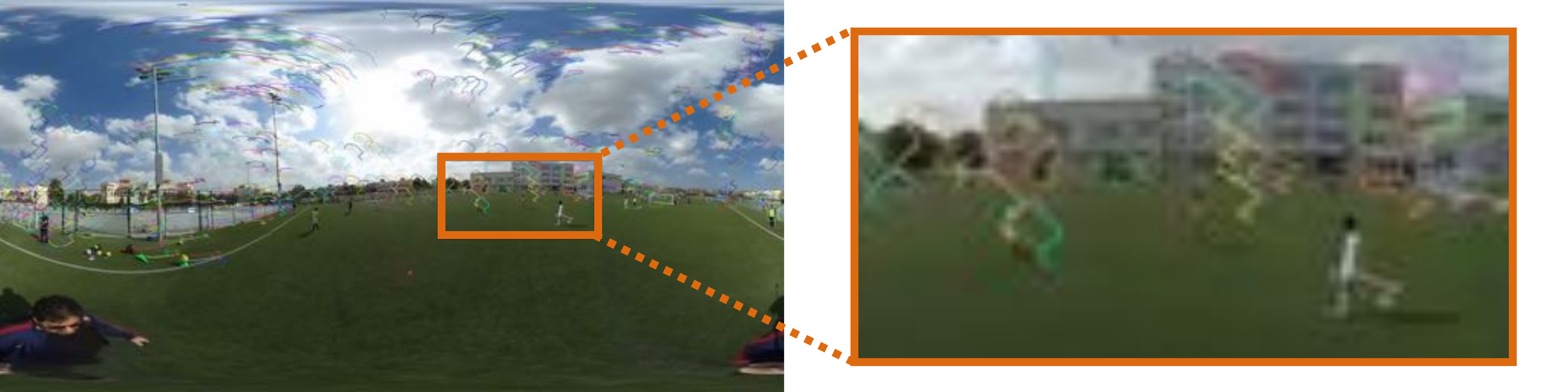}\\
		\begin{sideways}\hspace{2.5em}{Pano2Vid}\end{sideways}& 
		\includegraphics[width=\linewidth,page=2]{figures/wstabilization.pdf}\\ 
		\begin{sideways}\hspace{2.5em}{Ours}\end{sideways}& 
		\includegraphics[width=\linewidth,page=3]{figures/wstabilization.pdf} 
	\end{tabular}
	\caption{\label{fig:wostabilization}\tog{The feature trajectories of the input and Pano2vid~\cite{supano2vid} are shaky while our result is more stable and maintains the direction constraint from pano2vid.}}
\end{figure}

\paragraph{Direction w/o Stabilization.} 
\tog{Some recent works~\cite{supano2vid,DBLP:journals/corr/SuG17,HuLLCCS17} focus on converting spherical videos to traditional narrow FoV videos. 
These works re-direct the video, but do not consider the camera motion in the original videos, assuming that the input is captured by static or smooth cameras.
This assumption does not hold for many consumer videos captured by hand-held cameras with shaky movement. 
As shown in Fig.~\ref{fig:wostabilization}, feature trajectories in the output of Pano2Vid are as shaky as the input, while our result not only focus on the important content but achieves this with a smooth camera path.}


\begin{figure}
	\begin{tabular}{*{3}{c@{\hspace{3px}}}}
		\begin{sideways}\hspace{2.8em}Input\end{sideways}& 
		\includegraphics[width=.45\linewidth]{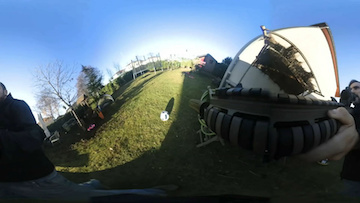} &
		\includegraphics[width=.45\linewidth]{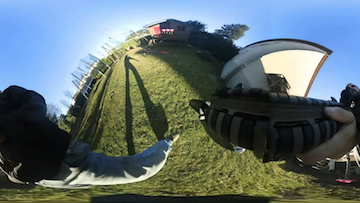} \\
		\begin{sideways}\hspace{2.5em}{Directed}\end{sideways}& 
		\includegraphics[width=.45\linewidth]{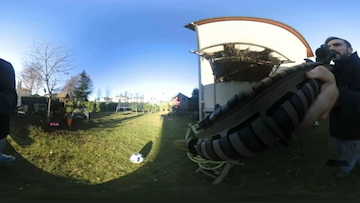} &
		\includegraphics[width=.45\linewidth]{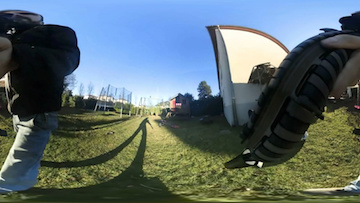} \\
		& Frame 0 & Frame 200\\
	\end{tabular}
	\caption{\label{fig:results2}In the input video (top) the camera is mounted on a toy gun, which is pointed down at the ground frequently . Using manual editing constraints we can keep the video pointed ahead even when the gun is lowered, making it much easier to watch.}
\end{figure}

\begin{figure*}[t]
	\centering
		\begin{tabular}{*{4}{c@{\hspace{3px}}}}	
		\includegraphics[height=2.3cm]{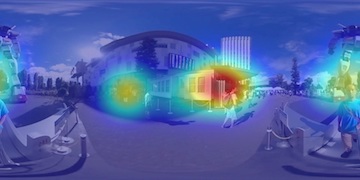}   & 
		\includegraphics[height=2.3cm]{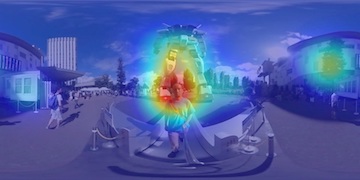} &
		\includegraphics[height=2.3cm]{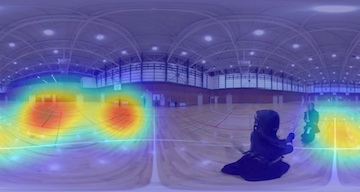}   & 
		\includegraphics[height=2.3cm]{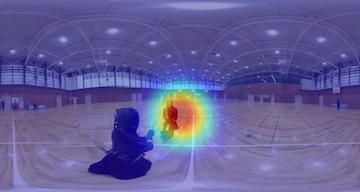} \\
		(a) Input & (a) Directed & (c) Stabilized & (d) Directed \\
			\end{tabular}
	\caption{\label{fig:userstudy}Gaze directions from our user study visualized as a heatmap over the video frames. In (a), a large portion of viewers will miss seeing the robot from the front, as they remain looking around the direction of motion. In (b), the camera rotates quickly, moving the important action out of view. It takes some time for users to find the kendo player again after this, whereas in the directed version, they can watch the whole scene without getting lost.}
\end{figure*}

\paragraph{Stabilization w/o Direction.} 
\tog{Using our method we can also generate stabilized-only results, and we provide some examples of this in the supplemental material.
In general, videos that are entirely stabilized are hard to watch, as camera motion and objects of interest can drift away from true north, causing viewers to get lost.
This can be seen in Fig.~\ref{fig:userstudy} (c,d), where when viewed with stabilization but without direction (b), the opposing kendo player moves around the camera quickly, and many viewers loose track of the players, requiring some time for their gaze to catch up to the action. 
However, in the directed version (c), the opposing player is kept in the center, which makes viewing much easier.
It is also possible to reintroduce a smoothed version of the original viewing direction back into the video~\cite{kopf2016360}, however this is not sufficient in many cases, as the original camera direction may often times not be ideal as shown in Fig.~\ref{fig:results2}, and also can be seen in Fig.~\ref{fig:userstudy} (a,b).}

\tog{Finally, as observed in our user study, and collaborated by other perceptual studies~\cite{8269807}, many viewers tend to watch 360\deg content passively.
We can see that the average viewing direction is highly centered in the true north direction, independent of the video content. 
Therefore, it is crucial to direct interesting events to this region.}

\begin{figure}
	\subfigure[Joint Optimization]{\includegraphics[width=0.49\linewidth,page=1]{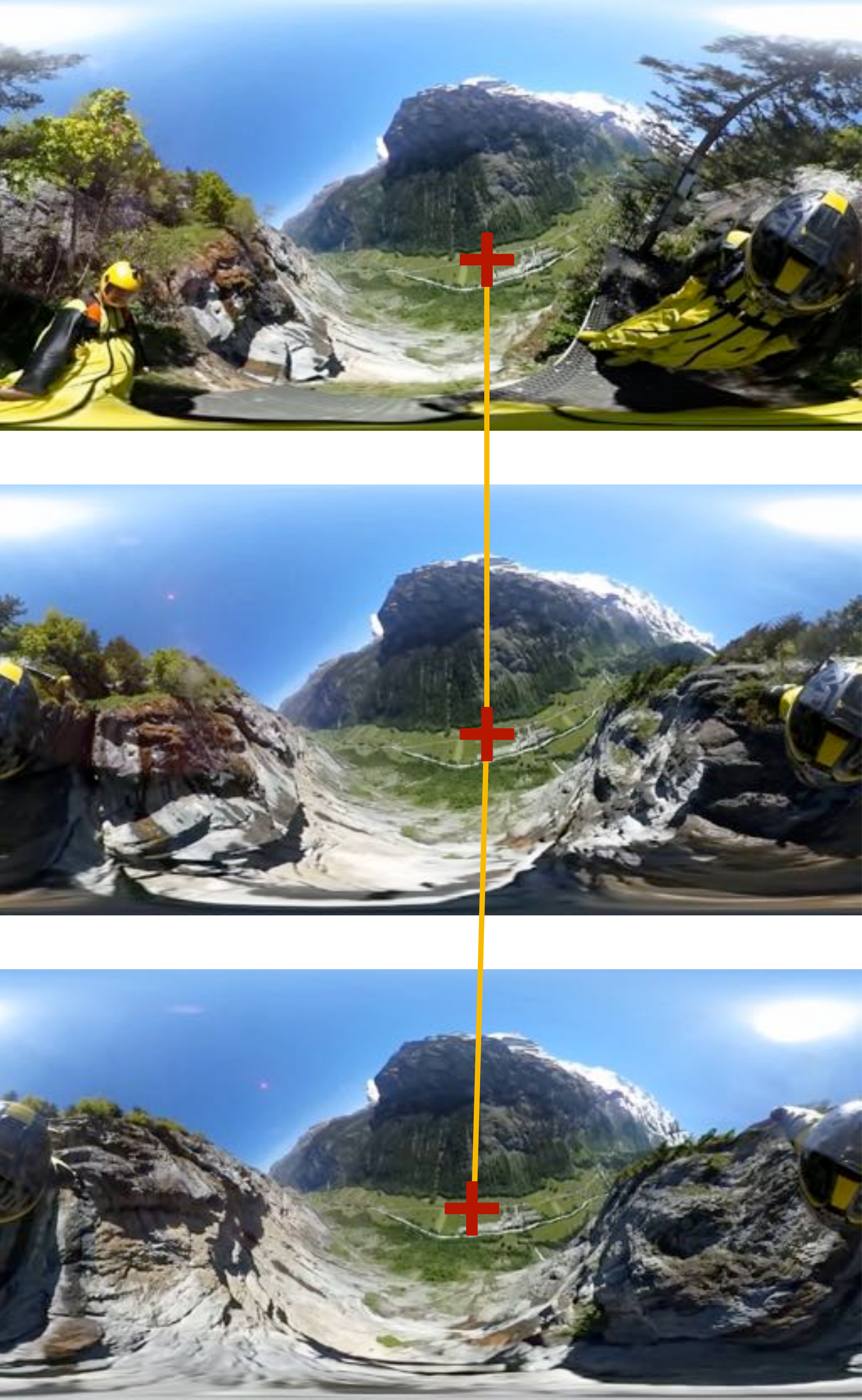}}
	\subfigure[Two-stage optimization]{\includegraphics[width=0.49\linewidth,page=2]{figures/jointvseperate.pdf}}
	\caption{\label{fig:jointoptimization}\tog{A two-stage optimization can not guarantee the direction in the output is consistent since the direction constraint is generated for each edited frame independently. So it is necessary to optimize with direction constraints and smoothness constraints jointly to regularize the direction constraint. }}
\end{figure}

\paragraph{Two-stage vs Joint Optimization.} 
\tog{
One question is whether it is necessary to optimize the direction and smoothness jointly, or if satisfactory results could be obtained by stabilizing the inputs first and then directing the output of stabilization. 
Some recent works~\cite{HuLLCCS17} uses the later strategy to generate 2D hyperlapse from 3D spherical videos. 
We compare the two-stage optimization with our joint approach by first stabilizing the video ignoring the direction constraint in Eq.~\ref{eq:cinema}, then we adjust the direction of the stabilized result by smooth interpolation of the positive direction constraints. 
As shown in Fig.~\ref{fig:jointoptimization}, our approach creates smoother camera paths overall, as direction constraints may be inconsistent across time, and being able to jointly solve for both gives us more flexibility to choose between multiple valid stable paths. 
This is especially true with automatically generated constraints such as forward motion or saliency. 
Please refer to the supplementary videos for more detailed comparison.
}







\subsection{User Study}

We additionally validate our approach by conducting a user study. 
\tog{
In this study, we attempt to answer two questions.
The first is whether our solution actually increases the chance that users will see the desired targets in the 360\deg video. 
Second, we attempt to qualitatively evaluate our results by determining user preferences of our results as compared to the inputs and fully stabilized versions.
}

\togb{
To do this, we recruited 20 users to participate in our study, with an age range from 25 to 50 years old.
Users were seated to simulate common viewing conditions for watching 360\deg video at home, and the videos were viewed using a Google Cardboard VR headset with an iPhone 6s.
We then conducted a two-alternative forced choice (2AFC) preference experiment, where viewers were presented with two versions of a video and after viewing both sequentially were asked which they preferred.
In addition, the gaze directions of each viewer was recorded, to measure the overlap with the positive constraint regions. 
The study included 6 videos, with an average length of 30 seconds. 
For each video, we compare the following versions: a) the original video, b) the stabilized only video, c) the directed and stabilized video via two-stage optimization and d) our directed and stabilized video via joint optimization.
Comparisons are made as pairwise choices, yielding 6 different pairwise comparisons per video. 
Within each pair, video positions are randomized, and the order of the 6$\times$6=36 pairs is drawn randomly.
Viewers were not given any specific viewing instructions besides simply exploring the videos as they like. 
}

\begin{table}[t]
	\centering
	\definecolor{rowblue}{RGB}{220,230,240}
	\setlength{\tabcolsep}{5pt} 
	\setlength\extrarowheight{1.8pt}
	\small
	\rowcolors{3}{rowblue}{white}
	\resizebox{.9\linewidth}{!}{	
		\begin{tabular}{lccc}
			\toprule
			& Original Video & Ours & Stabilized \\
			\midrule
			\midrule
			Mean distance (deg) & 57.89 & \textbf{25.87} & 76.69 \\
		
			Percent seen (\%) & 34.79 & \textbf{56.18} & 13.23 \\
			\bottomrule
		\end{tabular}
	}
	\vspace{.3cm}
	\normalsize
	\caption{Measured distance of participants' gaze directions to the positive directional constraints (lower is better). We also show the percent of positive constraints likely seen, which corresponds to the fraction of constraints that were within 30\deg of a viewer's gaze (higher is better).}
	\label{tbl:study}
\end{table}

To validate whether our method can be used to improve the viewing experience by directing the camera towards important events, we compute two quantitative measures using the recorded viewing tracks. 
First, we measure the mean distance of all users to all positive constraints in the videos. 
Second, we compute what percentage of viewers gaze was within 30\deg of the positive constraints.
We can see the results in Table~\ref{tbl:study}, indicating that adding direction to the camera significantly increased the chance that viewers will witness the events that the editor chooses to highlight.  
Figure~\ref{fig:userstudy} shows an example frame where the viewers have missed an important event in the undirected version.
This is an important step to validate, given that the main goal of our approach is to increase the chance of seeing positively marked events. 
This is mostly due to the inherent preference for forward-facing viewing, and the problem with getting lost when camera motion is not smooth. 
Please see the supplemental material for a visualization of viewing direction of participants.

\begin{figure*}[t]
	\centering
	\includegraphics[width=\linewidth]{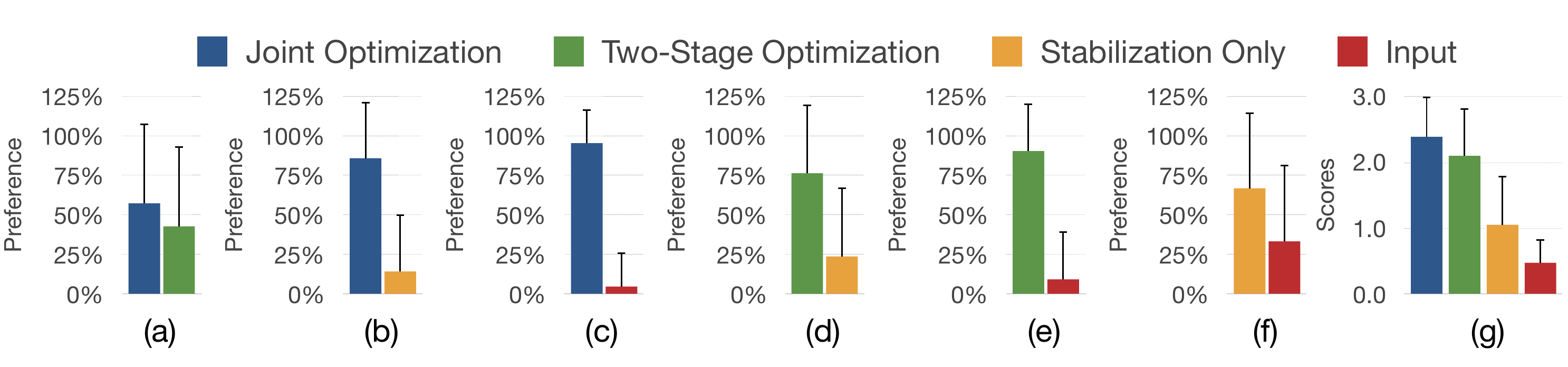} \\
	\caption{\togb{Percentage of obtained votes from the user study, showing a slight preference for (a) joint direction and stabilization v.s. two-stage direction and stabilization, and strong preferences for (b) joint direction and stabilization v.s. stabilization only, (c) stabilization only v.s. input, (d) two-stage direction and stabilization v.s. stabilization only, and (e) two-stage direction and stabilization v.s. input.  We also show a slight preference of (f) stabilization only v.s. input, and (g) the mean scores for each version.}}
	\label{fig:vote}
\end{figure*}

\togb{
In the second experiment, we perform a qualitative test, where users were asked for each pair, ``which video did you prefer to watch?''.
The results of this study are shown in Fig.~\ref{fig:vote}. 
We use one-way analysis of variance (ANOVA)~\cite{DixFinlayEtAl03} to validate the statistical significance of our results. 
To analyze the multiple preference test in a single ANOVA pass, we count the times that a specific version was selected as the preferred video, and use this as a score.
After collecting scores for all four different versions on all videos, we run one-way ANOVA to evaluate the statistics significance of the differences between different models.
We find that the p-value of the ANOVA is $7.269\times10^{-14}$, indicating that the conclusions of our subjective evaluation are reasonable. 
As shown in Fig.~\ref{fig:vote}, when compared with the input, a vast majority of users preferred our directed and stabilized version to both the input and stabilized footage. 
When compared within the directed and stabilized version, the joint optimization approaches is slightly preferred over the two-stage optimization, indicating that the joint optimization has found a good balance between stability and direction, although this result is less statistically significant. 
We also conduct a comparison between the stabilized version and the input, and found a slight preference of the stabilized version. 
}


\togb{
We note that we conducted a relatively long term user study, lasting 36 minutes for each viewer. 
To avoid viewers exceeding the maximum time that they were comfortable viewing 360\degree video, we split the 36 comparison pairs into 3 groups and let the viewers watch one group each time with rests in between.
Longer experiments would be needed to measure differences in viewing comfort over extended viewing sessions. 
}

\subsection{Limitations}
\tog{
Our method directs and stabilizes 360\deg or wide-angle videos based on estimated 3D camera motion and directional constraints. 
While this approach works well in many cases, it has some limitations. 
For one, we found that when videos were captured using multi-camera setups with imprecise temporal synchronization, the 3D scene assumptions in Sec.~\ref{sec:motionestimation} are no longer valid, and our method is unable to compensate the shakiness in the input video.
}

\tog{
Second, the joint direction and stabilization optimization in Sec.~\ref{sec:motionestimation} assumes that the input direction satisfies the cinematography rules. 
When the input direction constraint violates cinematographic rules, our joint optimization can not correct the direction constraint by smoothness.
We can see examples of this when viewing constraints are derived for example from automatic saliency computations in the supplemental material.
Additionally, when foreground objects, especially those with coherent motion tracks, occupy the majority of the viewing sphere, our optimization is unable to identify only the correct background tracks. 
Please see the supplemental video for examples of these cases.
}



%
\section{Conclusion and Future Work}
\togb{
In conclusion, we have presented an approach for joint stabilization and direction of 360\deg video. 
360\deg video is particularly well suited for this task, as all views are present at capture time, allowing full control over viewing direction in post.
In our work, we address modifying only the look-at direction, however this is just one small part of computational cinematography for 360\deg video, and there are many interesting areas for follow up work. 
For example, we do not experiment with changing focal length (zoom), as current 360\deg cameras do not have suitable resolution for close-up shots.
Recent work~\cite{serrano2017movie} has studied how people react to cuts in 360\deg viewing, especially when important content regions are inconsistent across the cut.
We validate our approach with a similar experiment, in that we look at the percentage of fixation points of viewers inside regions of interest over short videos.
However, our study is complementary, and shows how direction affects the viewing experience. 
These findings suggest our method could be used in conjunction with observations from Serrano et al.~\shortcite{serrano2017movie} to automatically direct content to be consistent over cuts. 
Finally, we believe that virtual cinema experiences with wide angle footage is a good way to bridge the gap between the wide availability of 360\deg viewing devices and the limited library of content.
To this end, our approach can be used with common wide angle first person cameras, as well as the recently introduced family of 180\deg VR cameras.  
}




\bibliographystyle{ACM-Reference-Format}
\bibliography{360stabilization}
\begin{appendix}
\label{appendix:a}
\section{Derivatives of Motion Estimation}
In this appendix, we give the analytic derivatives of our motion estimation.
For the convenience of representation, we denote $\ve{t}_{\times}\ve{R}\ve{p}$ in Eq.~\ref{eq:motionerror} as $\textbf{n}$ and multiply the angle $\omega$ to the rotation axis $\textbf{r}$ as $\textbf{w}$. We first give $\frac{\partial E}{\partial \textbf{n}}$, $\frac{\partial\textbf{n}}{\partial\textbf{w}}$ and $\frac{\partial\textbf{n}}{\partial\ve{t}}$, and then $\frac{\partial E}{\partial\textbf{w}}$ and $\frac{\partial E}{\partial\ve{t}}$ can be calculated using chain rule.

$\frac{\partial E}{\partial \ve{n}}$ is derived as:
\begin{equation}
\frac{\partial E}{\partial \textbf{n}}=\frac{\ve{q}^{\top}}{\sqrt{1-\big({\frac{\ve{q}^{\top}\ve{n}}{\|\ve{n}\|}}\big)^2}}\big(\frac{\ve{I}}{\|\ve{n}\|}-\frac{\ve{n}\ve{n}^{\top}}{\|\ve{n}\|^3}\big),
\end{equation}
where $\ve{I}$ is a $3\times3$ identity matrix.

$\frac{\partial\ve{n}}{\partial\ve{w}}$ and $\frac{\partial\ve{n}}{\partial\ve{t}}$ are:
\begin{equation}
\frac{\partial\ve{n}}{\partial\ve{w}}=\ve{t}_{\times}{(\ve{R}\ve{p})}_{\times},
\end{equation}

\begin{equation}
\frac{\partial\ve{n}}{\partial\ve{t}}={(\ve{R}\ve{p})}_{\times}.
\end{equation}

By chain rule we can finally get $\frac{\partial E}{\partial\ve{w}}$ and $\frac{\partial E}{\partial\ve{t}}$ as $\frac{\partial E}{\partial\ve{w}}=\frac{\partial E}{\partial \ve{n}}\frac{\partial\ve{n}}{\partial\ve{w}}$ and $\frac{\partial E}{\partial\ve{t}}=\frac{\partial E}{\partial \ve{n}}\frac{\partial\ve{n}}{\partial\ve{t}}$. 

\end{appendix}

\end{document}